\documentclass[twocolumn,showpacs,superscriptaddress,floatfix]{revtex4}

\usepackage{color}
\usepackage{tabularx}
\usepackage{epsfig}
\usepackage{amsmath}
\usepackage{amssymb}
\usepackage{graphicx}
\usepackage{bm}
\usepackage{psfrag}
\usepackage{enumitem}
\usepackage{times}
\usepackage{CJKutf8}

\usepackage{microtype}
\usepackage[svgnames]{xcolor}
\usepackage[pdftex]{hyperref}
\hypersetup{colorlinks=true,linkcolor=blue,citecolor=ForestGreen,urlcolor=DarkOrchid}

\begin{document}

\title{Feeding the multitude: A polynomial-time algorithm to improve sampling}

\author{Andrew J.~Ochoa}
\author{Darryl C.~Jacob}
\affiliation{Department of Physics and Astronomy, Texas A\&M University,
College Station, Texas 77843-4242, USA}

\author{Salvatore Mandr{\`a}}
\affiliation{Quantum Artificial Intelligence Laboratory, NASA Ames 
Research Center, Moffett Field, California 94035, USA}
\affiliation{Stinger Ghaffarian Technologies Inc., 7701 Greenbelt Road, 
Greenbelt, Maryland 20770, USA}

\author{Helmut G. Katzgraber}
\affiliation{Microsoft Quantum, Microsoft, Redmond, Washington 98052, USA}
\affiliation{Department of Physics and Astronomy, Texas A\&M University, College Station, Texas 77843-4242, USA}
\affiliation{Santa Fe Institute, 1399 Hyde Park Road, Santa Fe, New Mexico 87501 USA}

\date{\today}

\begin{abstract}

A wide variety of optimization techniques, both exact and heuristic,
tend to be biased samplers. This means that when attempting to find
multiple uncorrelated solutions of a degenerate Boolean optimization
problem a subset of the solution space tends to be favored while, in the
worst case, some solutions can never be accessed by the algorithm used.
Here we present a simple post-processing technique that improves
sampling for any optimization approach, either quantum or classical.
More precisely, starting from a pool of a few optimal configurations,
the algorithm generates potentially new solutions via rejection-free
cluster updates at zero temperature. Although the method is not ergodic
and there is no guarantee that all the solutions can be found, fair
sampling is typically improved. We illustrate the effectiveness of our
method by improving the exponentially biased data produced by the D-Wave
2X quantum annealer [S.~Mandr{\`a} {\em{ et al.}},Phys.~Rev.~Lett.~{\bf 118}, 070502 (2017)], as well
as data from three-dimensional Ising spin glasses. As part of the study,
we also show that sampling is improved when sub-optimal states are
included and discuss sampling at a finite fixed temperature.

\end{abstract}

\pacs{75.50.Lk, 75.40.Mg, 05.50.+q, 64.60.-i}

\maketitle

\section{Introduction}

Optimization problems, even when restricted to discrete (binary)
optimization, appear in many scientific disciplines and industrial
applications and typically map directly onto spin-glass-like
Hamiltonians \cite{stein:13,lucas:14}.  These problems are
computationally hard to solve with current hardware, which has led to
the design and construction of special-purpose analog quantum
optimization machines, such as the D-Wave Systems Inc.~quantum annealer.
In parallel, the development of efficient classical optimization
techniques to study these NP-hard problems has experienced a renaissance
in recent years.  Although many optimization techniques exist for
solving such problems, generally the complexity is worse than polynomial
in the size of the input. Therefore, it is of much interest to develop
efficient approaches to more efficiently study these systems.

While exact optimization techniques that obtain the optimum of a cost
function with guarantees are always desirable, these can typically only
handle a small number of variables
\cite{van:06,montanaro:15,mandra:16c}. Even worse, for degenerate
optimization problems they often return a single configuration
minimizing the cost function \cite{juenger:01,pardella:08,liers:10}.
Therefore, heuristics are the tool of choice, at the cost of obtaining
the solution of an optimization problem with a finite success
probability. It is therefore of much interest to develop techniques that
increase the quality and variety of solutions delivered by heuristics,
as well as improve the sampling of the solution space for degenerate
optimization problems. In parallel, sampling of uncorrelated solutions
at finite energy, as is needed, for example, in machine learning
applications, is currently also of much importance
\cite{torlai:16a,levit:17,crawford:18}.

In this paper we present a simple polynomial-time algorithm that can
substantially improve the sampling of degenerate ground states for
Ising-type Hamiltonians by starting from a subset of (suboptimal)
configurations.  This is of much importance when estimating the
ground-state entropy in physical systems, counting problems in computer
science such as \#SAT and \#Knapsack
\cite{jerrum:86,gomes:08,gopalan:11}, and industrial applications such as
SAT filters \cite{weaver:14,douglass:15}.  More precisely, if the
starting pool for our method contains only ground-state configurations,
our proposed method can potentially increase the variety of minimizing
configurations at a cheaper cost than other (often very expensive)
means. If low-energy configurations are also included in the pool, the
proposed method can be used to find configurations with a lower energy
and increase the probability of success to find the ground state.
Finally, if a pool of configurations at a finite energy (or temperature)
is given, our approach can be used to generate new, often uncorrelated
configurations at similar energy (or temperature).  As part of our
analysis, here we demonstrate how fair sampling
\cite{moreno:03,matsuda:09,mandra:17} of states generated with
transverse-field quantum annealing
\cite{finnila:94,kadowaki:98,brooke:99,farhi:01,santoro:02,das:05,santoro:06,das:08,morita:08}
using a D-Wave 2X quantum annealer \cite{comment:bias} can be improved.

We emphasize that the presented approach does not solve
optimization problems in polynomial time, as this would mean that ${\rm
P} = {\rm NP}$, nor is it guaranteed to generate all
solutions to a degenerate problem.  Indeed, one has to first find a
subset of solutions to jump-start the process and this task could be
exponentially difficult.  Nevertheless, a considerable amount of
resources can be saved by generating a new set of solutions in
polynomial time without the burden of running, e.g., time-consuming
algorithms.

The paper is structured as follows. In Sec.~\ref{sec:algoapp} we present
the algorithm as well as discuss potential areas where the approach
could have considerable impact. Section \ref{sec:experiments} contains
comprehensive benchmark results for different optimization problems as
well as discusses the performance of the algorithm for nondegenerate
optimization problems and sampling at finite temperature. Finally,
Sec.~\ref{sec:conc} summarizes our findings and presents an outlook for
future directions.

\section{Algorithm and applications}
\label{sec:algoapp}

In this section we first outline the algorithm, as well as variations
on how to apply it to perform different tasks, followed by a selection
of possible applications across disciplines.

\subsection{Outline of the algorithm}
\label{sec:algo}

The proposed algorithm is designed to perform large variable rearrangements on
Ising-type Hamiltonians \cite{yeomans:92} (i.e., Boolean variables) with
frustration. From a physics perspective, the most paradigmatic model
systems ideally suited to be studied with this approach are spin glasses
\cite{binder:86,stein:13}. However, because a plethora of discrete
optimization problems can be mapped directly onto spin-glass-like
Hamiltonians \cite{hartmann:04,lucas:14}, the method finds wide
applicability across many disciplines.

Our polynomial-time algorithm is based on the cluster updates first
introduced by Houdayer \cite{houdayer:01} for two-dimensional spin
glasses and later generalized to systems of arbitrary topology by Zhu
{\em et al.}~\cite{zhu:15b}. In its original incarnation, these cluster
updates are combined with Monte Carlo sampling to ensure ergodicity and
parallel tempering updates \cite{geyer:91,hukushima:96,katzgraber:06a}
to either thermalize a system at a low, but finite temperature, or find
optima for spin-glass-like Hamiltonians \cite{moreno:03,mandra:16}.
Furthermore, the cluster updates can be used as accelerators for
other optimization techniques, such as simulated annealing
\cite{kirkpatrick:83}. Here we strip the Monte Carlo aspect from the
algorithm and apply the cluster updates to variable configurations.

A cluster update is performed in the following way: First, compute the
site overlap between two variable $s_i^{(1)}$ and $s_i^{(2)}$ at site
$i$ of two different configurations $\{s^{(1)}\}$ and $\{s^{(2)}\}$,
$q_i = s_i^{(1)} s_i^{(2)}$. This creates two domains in $q$ space with
sites that have either $q_i = 1$ or $q_i = -1$.  Clusters are defined
as the connected components of these domains. A randomly chosen site
with $q_i = -1$ is used as the seed for a cluster that is built by
adding all the connected variables in the $q_i = -1$ domain with
probability $1$.  This means the approach is rejection free.  When no
more sites can be added to the cluster, the variables in both
configurations that correspond to the cluster in the overlap space where
$q_i = -1$ are flipped with probability $1$, irrespective of their
orientation.

It is important to note that, by construction, the combined
change in energy $\Delta E$ (value of the cost function) is zero
\cite{houdayer:01}.  However, in general the energy of one configuration
will increase by an amount $\Delta E^{(1)}$, whereas the energy of the
other configuration will decrease by the corresponding amount, i.e.,
$\Delta E^{(2)} = - \Delta E^{(1)}$. For the particular case of
ground-state configurations, this means that $\Delta E^{(1)} = \Delta
E^{(2)} = 0$. As such, if two configurations are chosen from a pool of
optima of a degenerate optimization problem, the resulting
configurations will remain in the ground-state manifold.

Note that if clusters percolate (span the extent of the lattice) the
cluster update does not produce two new configurations
\cite{houdayer:01}.  However, in frustrated spin systems, it has been
demonstrated \cite{zhu:15b} that cluster percolation is suppressed at
low enough temperatures (energies) due to frustration. This means that
in the study of low-energy states percolation should not affect the
efficiency of the algorithm we outline below.

To generate new configurations out of a pool of existing configurations
obtained by using any arbitrary heuristic, the algorithm randomly pairs
two configurations from the pool, feeds them into the cluster update, and
adds the resulting unique configurations back to the pool, unless these
are already in the pool, in which case we increment the histogram.
Therefore, the algorithm can be summarized as follows.

\begin{enumerate}[leftmargin=*]

\item{Start with a pool of ${\mathcal C}$ configurations computed with
any simulational method.}

\item{Randomly select two configurations from the pool $\{s^{(n)}\} \in
{\mathcal C}$, $n = \{1,2\}$.}

\item{Perform a Houdayer cluster update step using configurations
$\{s^{(n)}\}$. If the resulting configurations $\{\tilde{s}^{(n)}\}
\not\in {\mathcal C}$, add these new and unique solutions to the pool. If $\{\tilde{s}^{(n)}\}
\in {\mathcal C}$, increase the count of this solution in the histogram.}

\item{Iterate as desired or until no new configurations can be found.}

\end{enumerate}

\noindent We emphasize that the algorithm is not ergodic. This means that
there is no guarantee that, for example, all solutions to a degenerate
optimization problem can be found. However, as illustrated in
Sec.~\ref{sec:experiments} below, the approach is quite efficient and for an
increasing number of variables and/or problem degeneracy performs increasingly
better while approaching a limiting distribution as discussed in the Appendix. Furthermore, the method can
be used at finite temperature by feeding a pool of finite-temperature
configurations to the cluster update. 

There are several stopping criteria one can implement. For example, one
can choose to stop when all solutions have been found, if they are
known. However, this is typically not the case. Another approach is to
stop when the empirical distribution of the accessible states is
comparable to a distribution obtained by selecting from a uniform
distribution (up to intrinsic statistical biases). While exhaustive
enumeration of solutions rather than randomized flipping of domains can
be a sensible approach, when applying this algorithm to random
instances, we do not know the size of the solution manifold before
starting the resampling.  In addition, exhaustive enumeration of
solutions does not scale for highly degenerate problems.

The success of the algorithm largely depends on the number of clusters
in the initial pool of solutions.  In this work the randomly generated
instances we have used to benchmark the resampling method can have very
different ground-state manifolds: some connected by a single spin flip
or clusters composed of a single spin, as well as other larger clusters
composed of multiple spins that equate to multiple spin flips.  Should the
initial set manifold consist of only spin-reversed ground states, the
algorithm will not find new solutions or escape this particular region
or regions of the ground-state manifold. In the case that there are more
solutions to be found, this can be alleviated by adding excited states
into the initial resampling pool. If such excited states are not readily
available, they are easily generated by randomly flipping a fraction of
variables in each ground-state configuration used in the initial
solution pool.

\subsection{Possible implementations}

Note that any two configurations can be fed into the cluster
update. Furthermore, the approach is easily parallelized on specialized
hardware. Below we list possible ways in which the algorithm can be
implemented with different goals in mind:

\begin{itemize}[leftmargin=0.0em]

\item[]{ (i)Expand the solution pool of a degenerate problem (ground
states only). Start from a subset of solutions to a degenerate
problem. Feeding the configurations to the cluster update could generate
by design new ground-state configurations (if possible).}

\item[]{ (ii) Expand the solution pool of a degenerate problem (include
low-energy states). Start from a pool of low-energy configurations
and feed these to the cluster update. If the ground-state energy is
known, keep track of all configurations that minimize the cost function.
Although some solutions will be ``lifted'' from the ground-state
manifold, having low-lying excited states vastly improves the sampling
of the ground-state manifold.}

\item[]{ (iii) Improve solutions of both degenerate and nondegenerate
optimization problems. Start from a pool of low-energy
configurations and feed these to the cluster update. Keep track of the
lowest cost-function values. In some cases, it is even possible to find
the ground state for nondegenerate problems \cite{comment:dorband}.}

\item[]{ (iv) Sample solutions at a fixed but finite temperature
(energy). Start from a pool of configurations computed at a given
finite temperature (or average energy per spin) and feed these to the
cluster algorithm. Keep track of the configurations whose energy falls
within a desired energy window. Add the new configurations to the pool.}

\end{itemize}

\noindent The aforementioned implementations illustrate a handful of
ways the heuristic can be used to sample states of discrete optimization
problems. We note that, quite often, the computational effort to
generate the initial pool of states is sizable, especially for NP-hard
problems, and as such, having a simple heuristic that can generate
new configurations in polynomial time could be transformative for a very
broad set of applications. In particular, for fully connected graphs,
the presented approach scales as $O(N^2)$, where $N$ is the
size of the input.

\subsection{Application scope}
\label{sec:app}

Because both ground-state and excited-state configurations can
be used and because the cluster update can generate new configurations
from existing states with potentially large Hamming distances if
existing states with large Hamming distances are used, the approach
finds wide applicability across disciplines. We list some application
domains below. However, we emphasize that the method can be applied to
any optimization problem where an overlap between two
configurations can be constructed and where the interactions are
symmetric, such as for a quadratic or higher-order unconstrained
optimization problem.

\begin{itemize}[leftmargin=0.0em]

\item[]{{\em (i) Improving data quality in, e.g., quantum annealing
machines.} Because quantum annealers such as the D-Wave Systems
Inc.~devices perform the optimization step several times to improve the
success probability of the solutions, a subset of obtained low-energy
configurations can be fed into the cluster update to generate
lower-energy solutions or thus increase success probabilities.}

\item[]{{\em (ii) Improving fair sampling for biased samples.} It has been
demonstrated \cite{matsuda:09,mandra:17} that transverse-field quantum
annealing is a biased sampling approach. Using the cluster update, the
solution pool can be expanded and the biased sampling mitigated.}

\item[]{{\em (iii) Fast generation of uncorrelated solutions for SAT
membership filters.} --- Probabilistic SAT membership filters
\cite{weaver:14} rely heavily on a large pool of uncorrelated solutions
to a complex SAT formula to reduce the filter's false-positive rate.
However, many SAT solvers tend to generate correlated solutions. Feeding
a pool of these to the cluster update can result in new solutions with
larger Hamming distances, therefore reducing correlations and thus the
false-positive rate.}

\item[]{{\em (iv) Training of machine learning techniques.} Machine
learning approaches, such as general or restricted Boltzmann machines,
require (ideally uncorrelated) training sets by sampling from a
thermalized system at a user-defined temperature. By feeding training set
configurations to the cluster update combined with post-selection,
uncorrelated solutions can be readily obtained to better train the
system. This is of much importance for quantum implementations that
use quantum annealing with a transverse-field driver
\cite{benedetti:16,levit:17,crawford:18}.}

\item[]{{\em (v) Chemistry applications, such as molecular similarity.}
Because the molecular similarity problem in chemistry can be cast as a
discrete optimization problem \cite{hernandez:17}, finding solutions
from a limited set can be improved by feeding the configurations to the
cluster update.}

\item[]{{\em (vi) Finance applications.} When searching for arbitrage
opportunities, speculators might not necessarily be interested only in
the optimal opportunity. By casting the problem in a quadratic
unconstrained binary optimization problem \cite{rosenberg:16x}, one can
use standard optimization techniques to find optima. Feeding these in
conjunction with any low-energy states to the cluster update might
result in alternate opportunities, thus helping the speculator diversify
the portfolio.}

\item[]{{\em (vii) Inclusion of constraints.} In an effort to simplify
optimization problems, certain constraints are typically neglected when
mapping the problem to a binary quadratic form. As an example, the
requirement that a traveling salesperson path is closed can initially be
neglected and later included as a constraint. A pool of minimizing
configurations can be fed to the cluster update to generate more
uncorrelated solutions that might better suit the additional
constraints.}

\item[]{{\em (viii) Acceleration of resampling techniques.} --- Sample
persistence optimization techniques \cite{karimi:17a} require a diverse
set of samples to work efficiently. By feeding the sample set to the
cluster algorithm, potentially further persistent samples can be
obtained, thus accelerating the optimization.}

\end{itemize}

\noindent In what follows we demonstrate the efficiency of the approach
using spin-glass Hamiltonians on both quasiplanar chimera graphs
where bipartite K$_{4,4}$ cells are connected on a squarelike lattice 
\cite{bunyk:14}, and three-dimensional topologies.

\section{Experiments}
\label{sec:experiments}

We demonstrate the efficiency of the cluster update heuristic on paradigmatic
optimization problems. The data sets analyzed either stem from numerical
simulations or were produced on the D-Wave 2X quantum annealer.

\subsection{Benchmark problem}
\label{sec:bench}

The simplest hard Boolean optimization problem is a spin glass
\cite{binder:86,stein:13}. The Hamiltonian (cost function) of a generic
spin-glass model is given by
\begin{equation}
{\mathcal H} = 	\sum_{\{i,j\} \in {\mathcal E}} J_{ij} S_i S_j 
		- \sum_{i \in {\mathcal V}} h_i S_i . 
\label{eq:ham}
\end{equation}
Here, $S_i \in \{\pm 1\}$ are Boolean Ising variables placed on vertices
${\mathcal V}$ of a graph ${\mathcal G}$ with edges ${\mathcal E}$. The
couplers $J_{ij}$ on the edges ${\mathcal E}$, as well as the biases
$h_i$ on the vertices ${\mathcal V}$, fully define the problem. In the
experiments performed below we set $h_i = 0$ $\forall i \in {\mathcal
V}$ without loss of generality. For the experiments we use different
graph topologies ${\mathcal G}$ as well as different coupler
distributions.

Configurations were obtained using the D-Wave Systems Inc.~D-Wave 2X
quantum annealer and simulated annealing \cite{isakov:15}. For the
three-dimensional lattices the Cologne spin-glass server
\cite{juenger:sg} was used to verify the optima.

\subsection{Quasi-planar chimera lattices}
\label{sec:chimera}

\begin{table}
\caption{
Number of disorder instances $N_{\rm s}$ from Mandr{\`a} {\em et~al.} 
\cite{mandra:17} sorted by system size $N$ and number of ground states 
$G = 3 \times 2^k$. For each system size and ground-state degeneracy,
the cluster update was applied $2^{20}$ times to the data set to produce
new states \cite{comment:cpu}.
\label{tab:mandra}
}
\begin{tabular*}{\columnwidth}{@{\extracolsep{\fill}} l c c c c c c r}
\hline
\hline
$N$ & $G\!=\!24$ & $G\!=\!48$ & $G\!=\!96$ & $G\!=\!192$ & $G\!=\!384$ &$G\!=\!768$ & $N_{\rm s}$\\
\hline
$512$ & $63$ & $51$ & $48$ & $26$ & $0$  & $0$  & $188$\\
$648$ & $70$ & $56$ & $59$ & $75$ & $0$  & $0$  & $260$\\
$800$ & $28$ & $52$ & $32$ & $59$ & $38$ & $6$  & $215$\\
$968$ & $22$ & $15$ & $31$ & $30$ & $28$ & $21$ & $147$\\
\hline
\hline
\end{tabular*}
\end{table}

We first demonstrate how the cluster updates can improve data sets
generated on the D-Wave 2X quantum annealer. We thus reanalyze the data
\cite{comment:params} from Ref.~\cite{mandra:17}. To perform a
systematic study, Ref.~\cite{mandra:17} selected the couplers in
Eq.~\eqref{eq:ham} from the Sidon set \cite{katzgraber:15,zhu:16}
$J_{ij} \in \{\pm 5, \pm 6, \pm 7\}$. For chimera lattices with $N = 8
\times c^2$ ($c = 4, \ldots, 12$, modulo broken qubits
\cite{comment:qubits}) sites, we only use ground states where the
degeneracy $G$ is $G = 3\times 2^k$ ($k \in {\mathbb N}$). We emphasize
that the chosen problem has a relatively small degeneracy $G$ compared
to other paradigmatic disorder distributions, such as bimodal couplers.
However, as we will illustrate below, the cluster approach becomes
increasingly efficient for larger values of $G$. Simulation parameters
are listed in Table ~\ref{tab:mandra}.

\begin{figure*}
\centering
\includegraphics[width=0.45\textwidth]{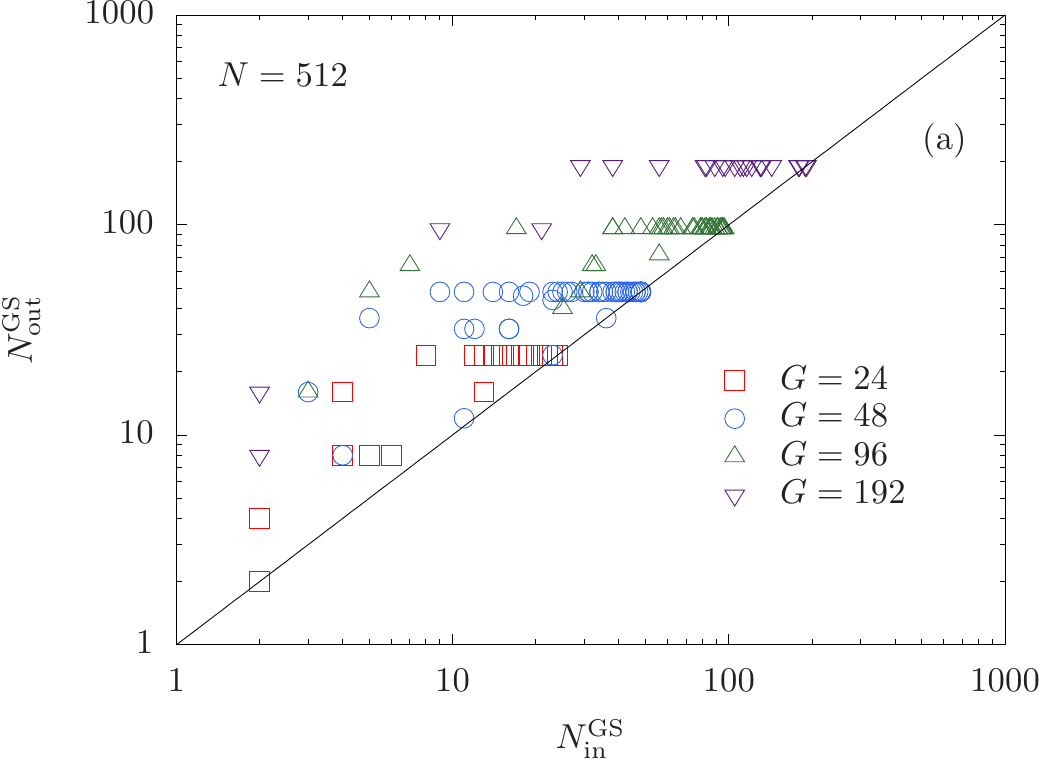}
\includegraphics[width=0.45\textwidth]{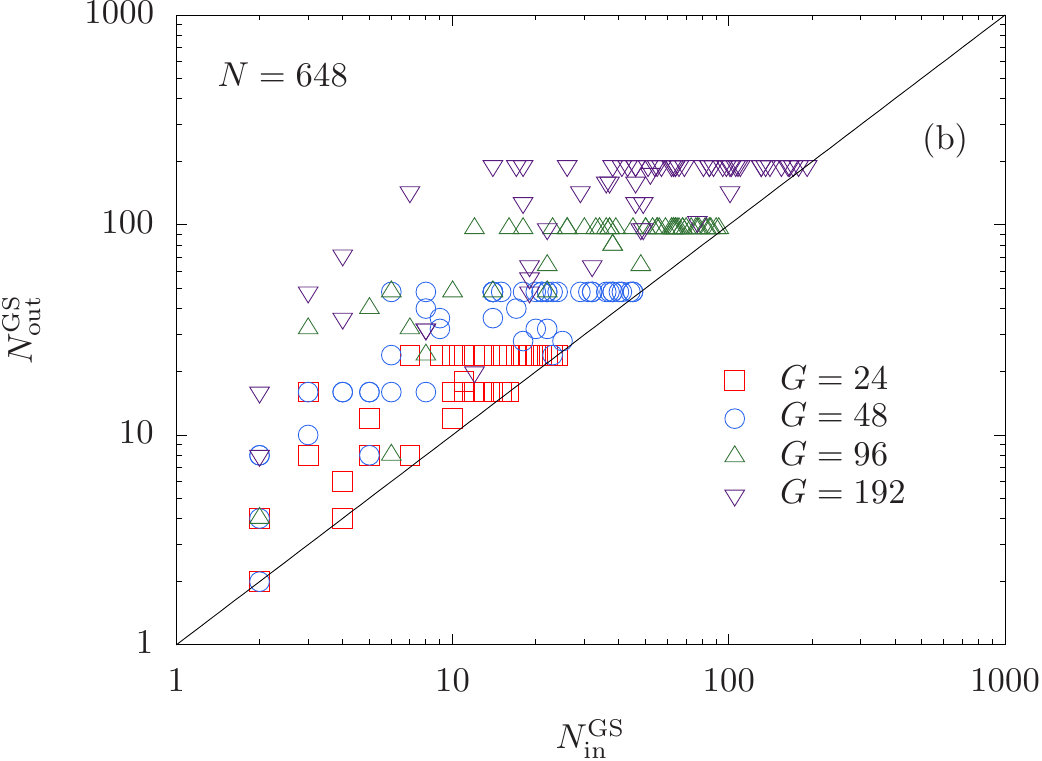}\\[1.5em]
\includegraphics[width=0.45\textwidth]{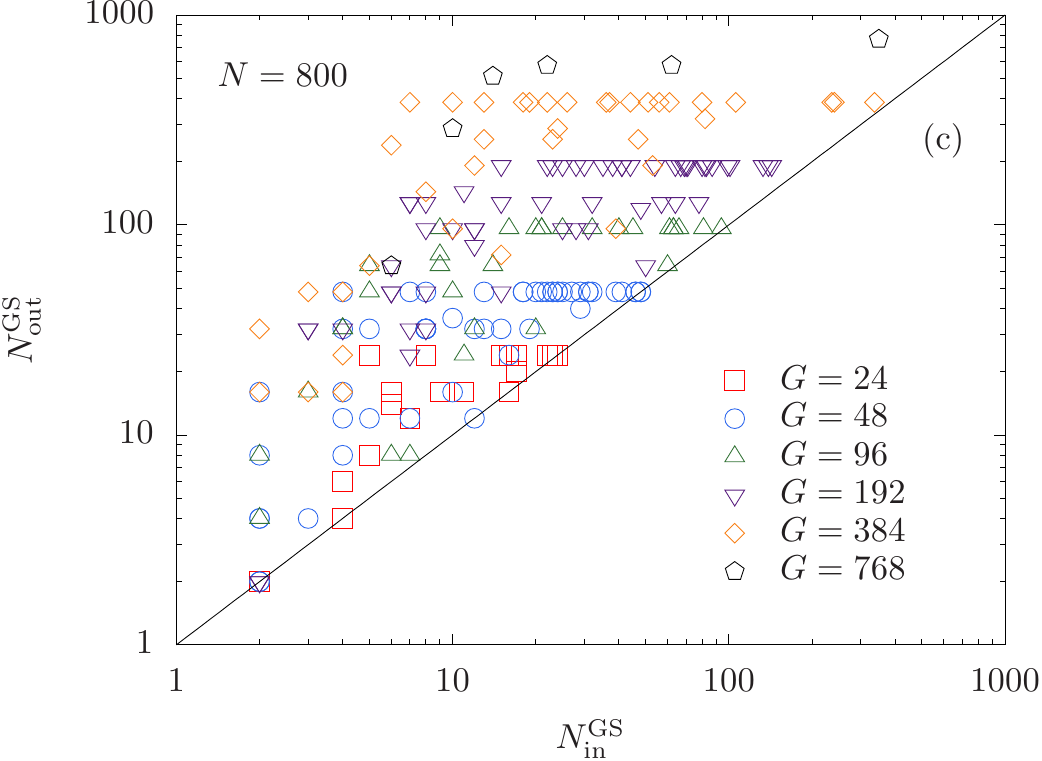}
\includegraphics[width=0.45\textwidth]{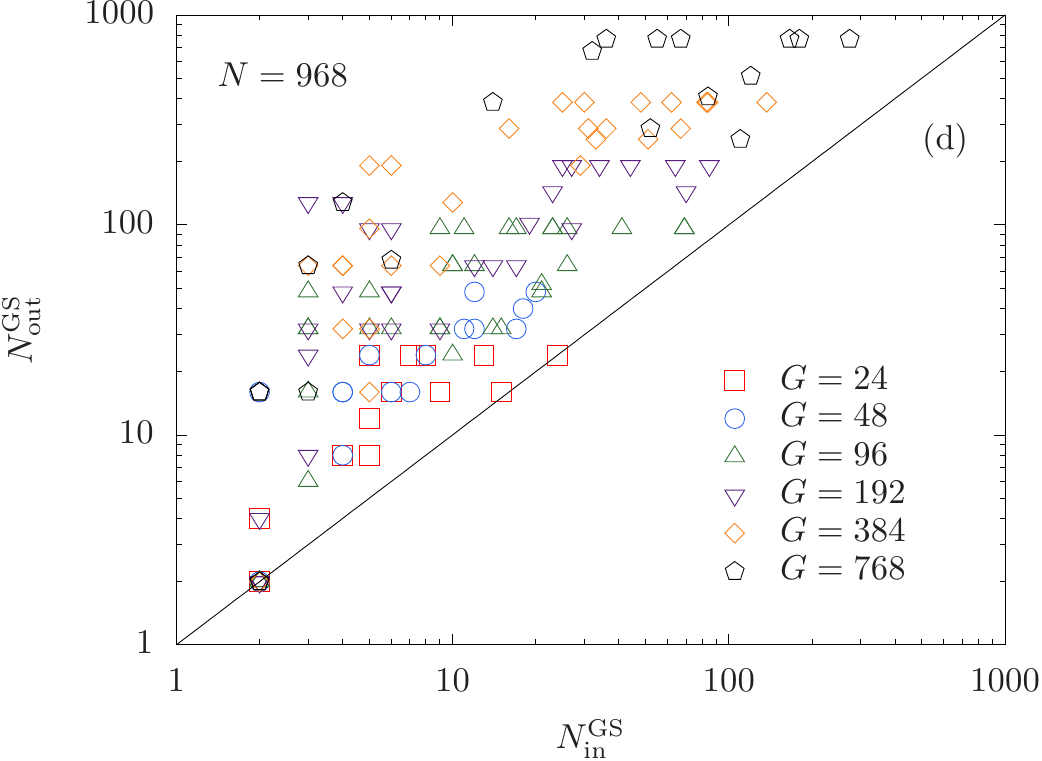}
\caption{
Number of solutions generated using cluster updates
($N^{\rm{GS}}_{\rm{out}}$) as a function of ground states found using the
D-Wave 2X quantum annealer ($N^{\rm{GS}}_{\rm{in}}$) for
different system sizes $N$ and known degeneracies $G$: (a) $N = 512$, (b) $N =
648$, (c) $N = 800$, and (d) $N = 968$. Each point
represents an individual instance.  Any points to the left of the
$N^{\rm{GS}}_{\rm{out}} = N^{\rm{GS}}_{\rm{in}}$ (solid diagonal line)
mean that additional minimizing configurations are found. For points
that lie on the line, either all solutions were already known or the
cluster updates produced no improvement.
}
\label{fig:C_GS}
\end{figure*}

\begin{figure}
\centering
\includegraphics[width=0.45\textwidth]{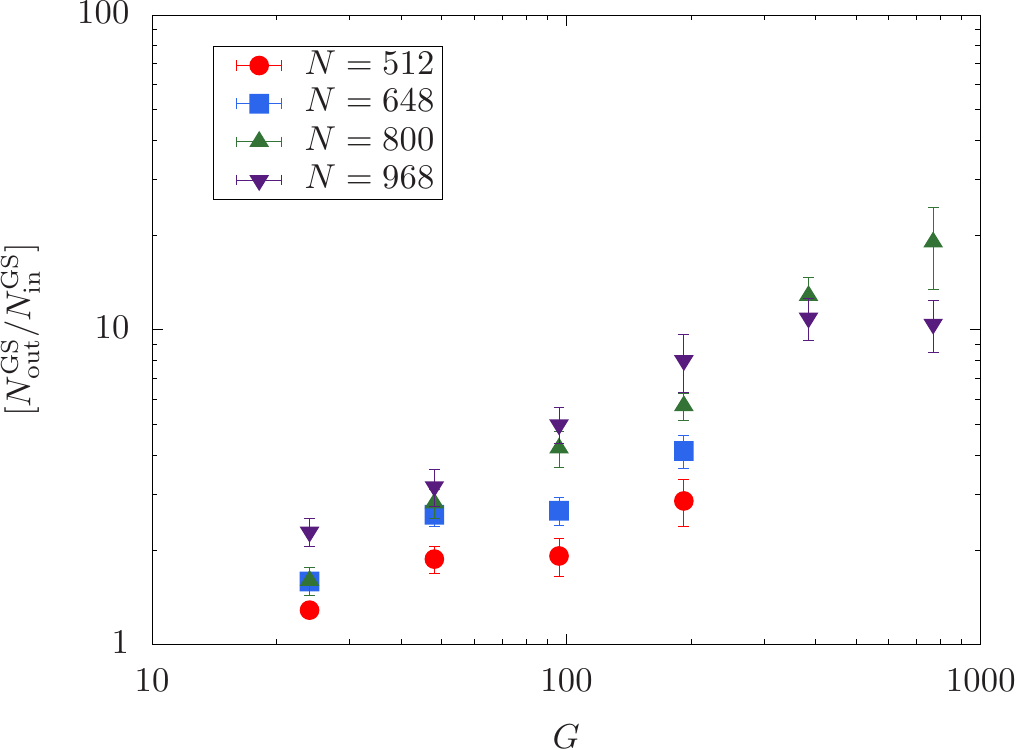}
\caption{
Relative improvement $[N^{\rm{GS}}_{\rm{out}} / N^{\rm{GS}}_{\rm{in}}]$
averaged over instances as a function of known degeneracy $G$ for
different system sizes $N$ after postprocessing the minimizing
configurations obtained with the D-Wave 2X quantum annealer.  Square
brackets represent a disorder average over all instances.
}
\label{fig:C_GS_i}
\end{figure}

Figure \ref{fig:C_GS} shows scatter plots of individual instances whose
minimizing configuration was obtained on the D-Wave 2X quantum annealer
for different system sizes $N$ and ground-state degeneracies $G$.
 Here $N^{\rm{GS}}_{\rm{in}}$ minimizing configurations are fed into the
algorithm (horizontal axis) and $N^{\rm{GS}}_{\rm{out}}$ are produced
after the post processing (vertical axis). The data show that for most
instances additional solutions are found (points left of the diagonal
line) using the cluster updates and that the effect is more pronounced
for larger degeneracies $G$.  For points that lie on the line, either all
solutions were already known or the cluster updates produced no
improvement. To better quantify the improvement over the original D-Wave
results, we study the disorder-averaged ratio $[N^{\rm{GS}}_{\rm{out}} /
N^{\rm{GS}}_{\rm{in}}]$ as a function of the degeneracy $G$ for
different system sizes $N$ in Fig.~\ref{fig:C_GS_i}. For both increasing
$N$ and $G$ the cluster updates perform increasingly better. We do
emphasize, however, that for small $N$ and $G$ most if not all
minimizing configurations are found by the D-Wave 2X, whereas for
large $N$ and $G$ this is less probable. Therefore, there could be an
intrinsic bias in the way we present the data. However, it is clear that
our post-processing of the data vastly improves the solution
space generated using the quantum annealer.

We now expand the data set produced by the D-Wave 2X quantum annealer in
Ref.~\cite{mandra:17} by including first-excited states in the
resampling. When the algorithm is run, states that minimize the cost
function are recorded while any other produced states are kept in the
pool.  This results in a clear advantage over the sampling of ground
states: As mentioned, the cluster updates are not ergodic. This means
that if the subset of minimizing configurations is small or from the
same region in phase space, it will be hard for the algorithm to find
other configurations in ``more remote'' parts of the phase space.
However, by allowing first-excited states, this problem can be partially
overcome. While some nonminimizing states are pushed into the
ground-state manifold thus enriching the solution pool, some ground
states are lifted out of the ground-state manifold into excited states
to be pushed back at a later stage of the heuristic, however into a
different part of phase space. For these experiments the cluster updates
were applied $2^{17}$ times to each data set.

We note that the study can in principle be repeated for any number of
low-lying states and does not need to be restricted to the first-excited
state. In fact, the inclusion of higher-energy levels will allow the
algorithm to more efficiently sample phase space.

\begin{figure*}
\centering
\includegraphics[width=0.45\textwidth]{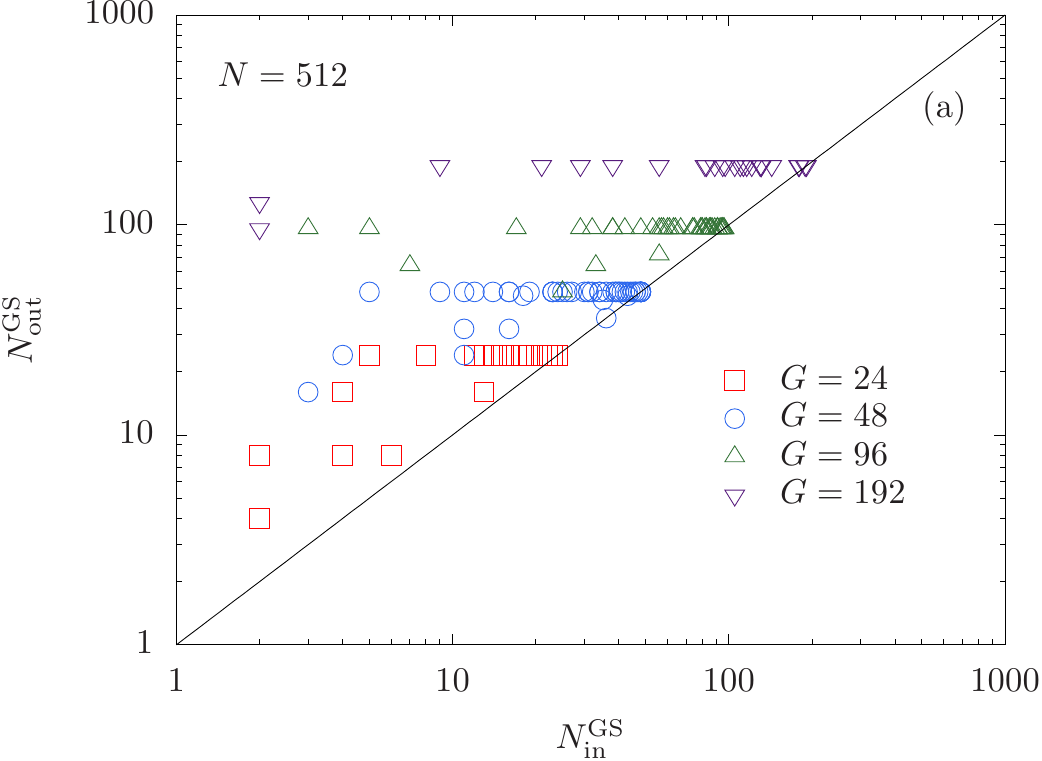}
\includegraphics[width=0.45\textwidth]{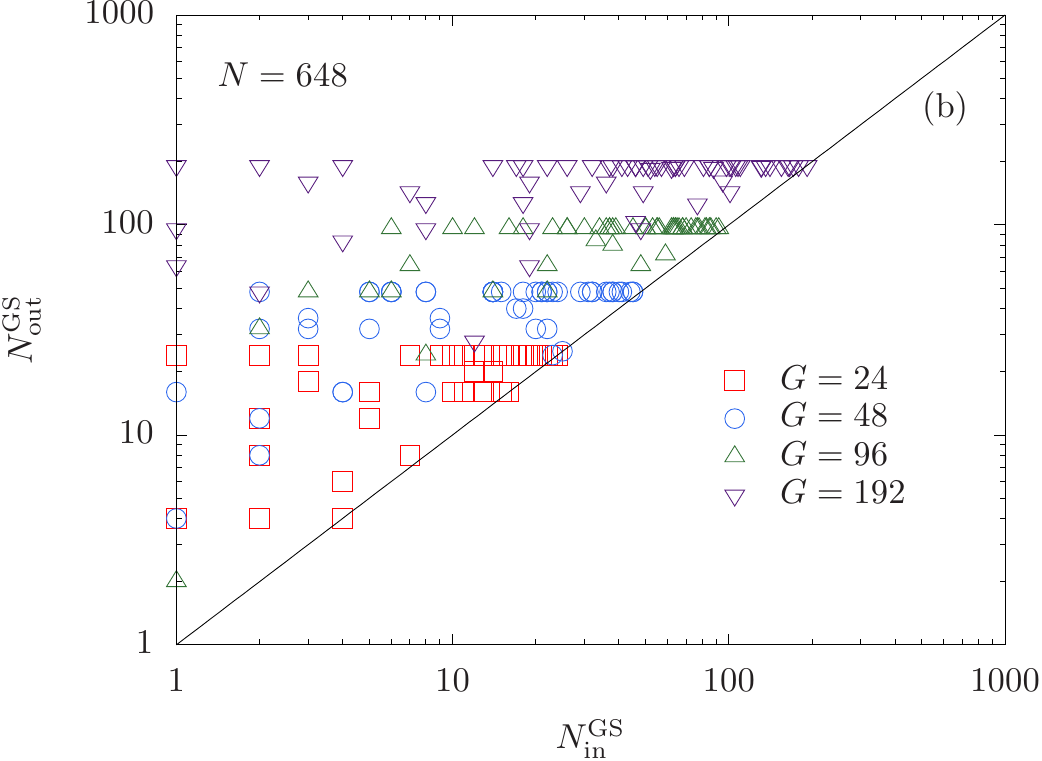}\\[1.5em]
\includegraphics[width=0.45\textwidth]{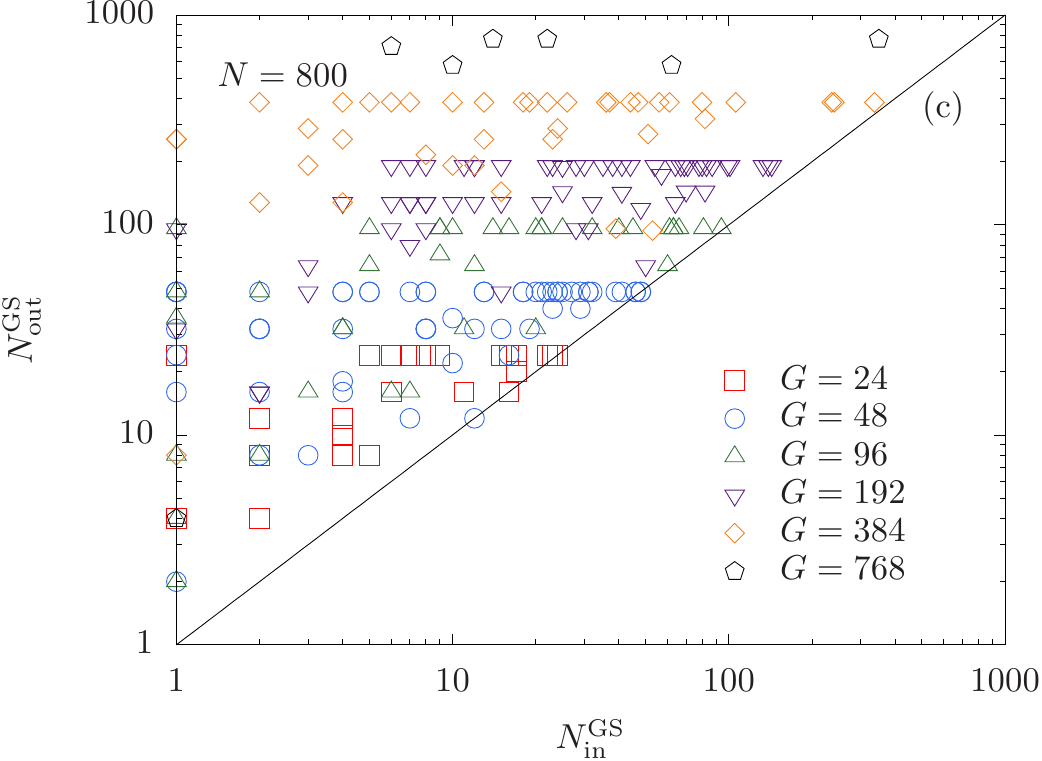}
\includegraphics[width=0.45\textwidth]{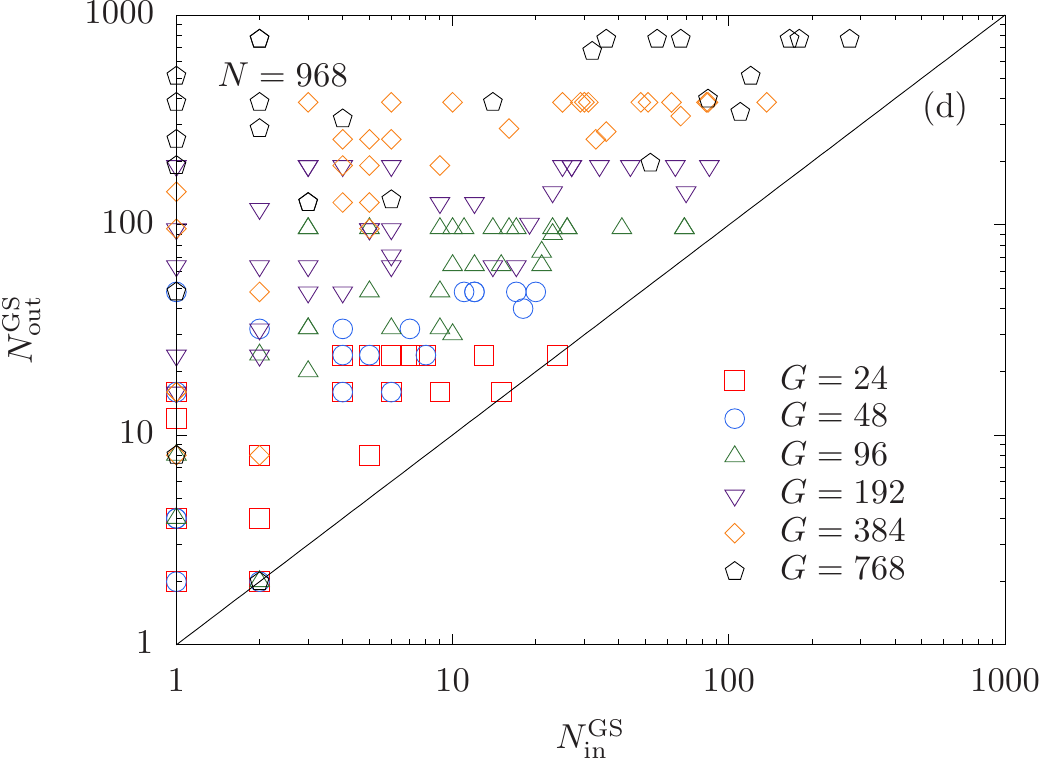}
\caption{
Number of solutions generated using cluster updates
($N^{\rm{GS}}_{\rm{out}}$) as a function of ground states found using
the D-Wave 2X quantum annealer ($N^{\rm{GS}}_{\rm{in}}$) for different
system sizes $N$ and known degeneracies $G$: (a) $N = 512$, (b) $N =
648$, (c) $N = 800$, and (d) $N = 968$. Each point represents an
individual instance.  Note that both ground states and first-excited states were fed into the cluster update in this experiment.
Including a more diverse set of configurations and keeping track of
minimizing configurations results in a better sampling of the
ground-state manifold, as can be seen for cases where only one
ground-state configuration was available and many more ground states
were obtained by adding first excited states into the state pool (data 
points on the vertical axis).
}
\label{fig:C_GS_ES}
\end{figure*}

\begin{figure*}
\centering
\includegraphics[width=0.45\textwidth]{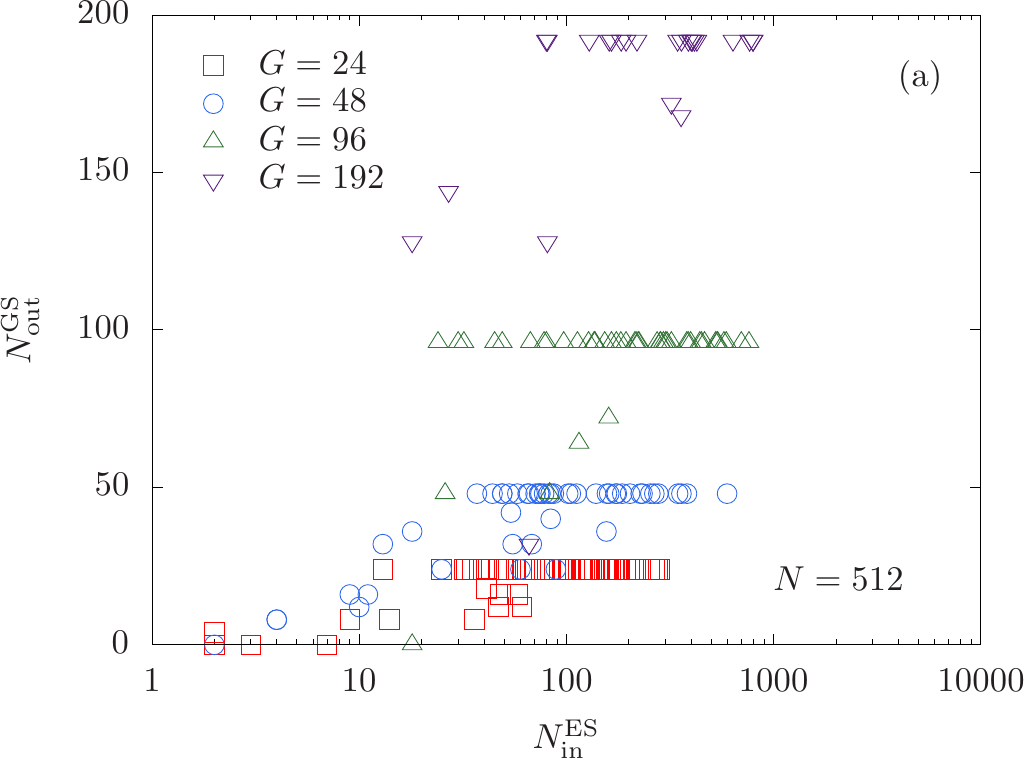}
\includegraphics[width=0.45\textwidth]{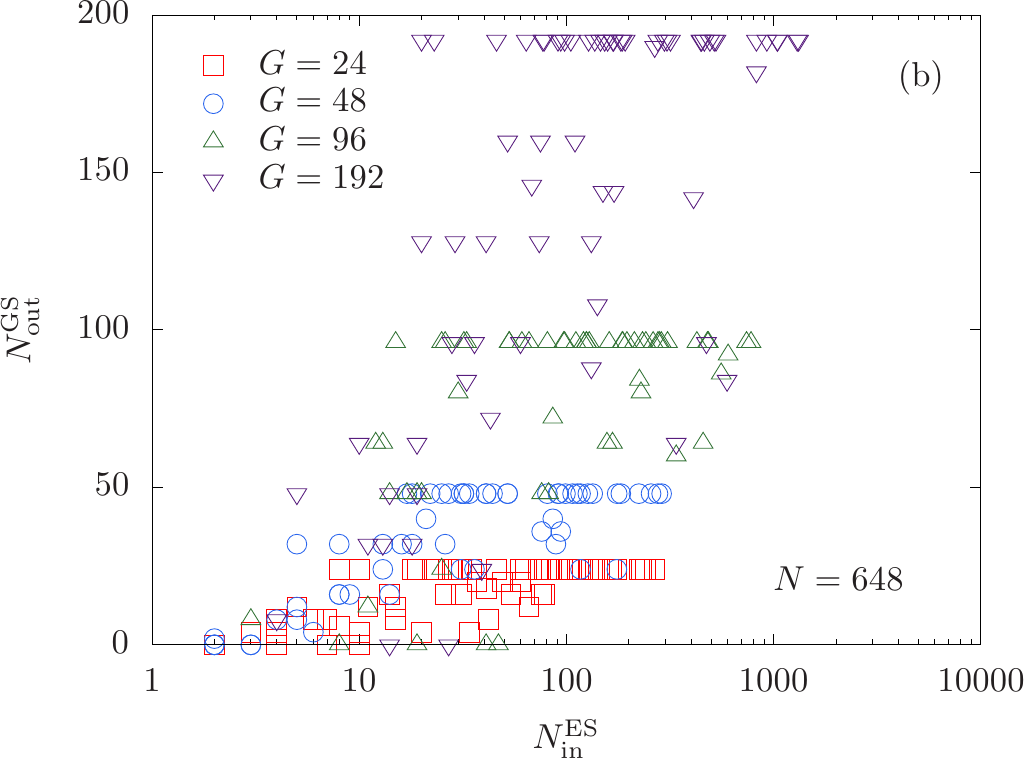}\\[1.5em]
\includegraphics[width=0.45\textwidth]{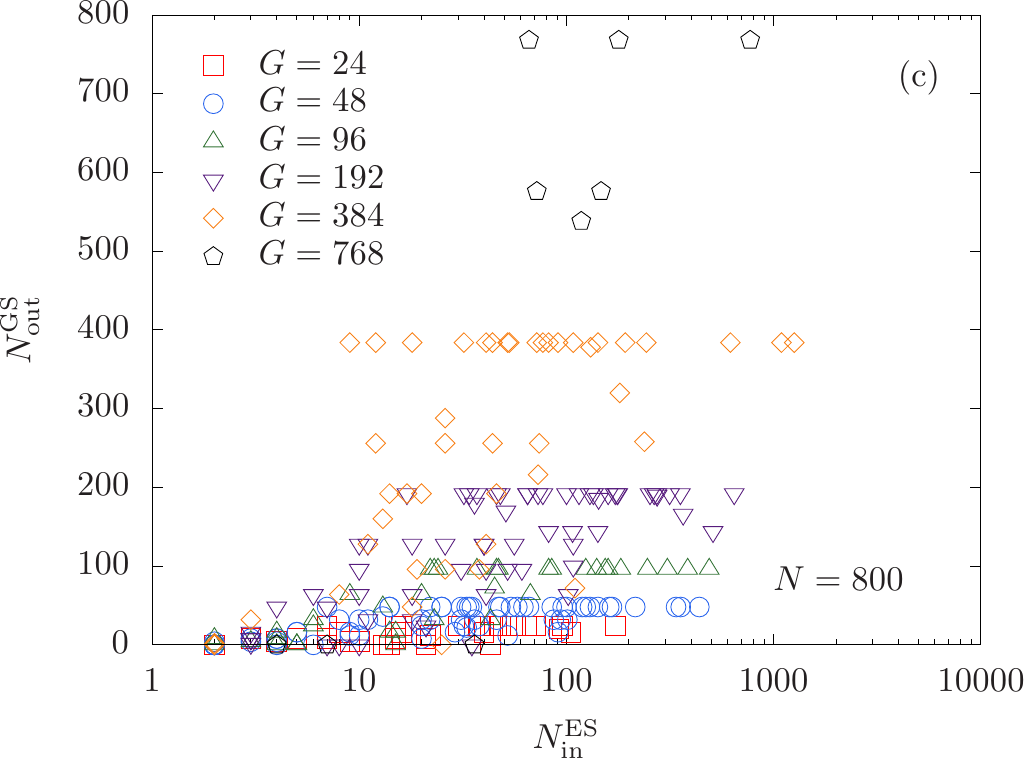}
\includegraphics[width=0.45\textwidth]{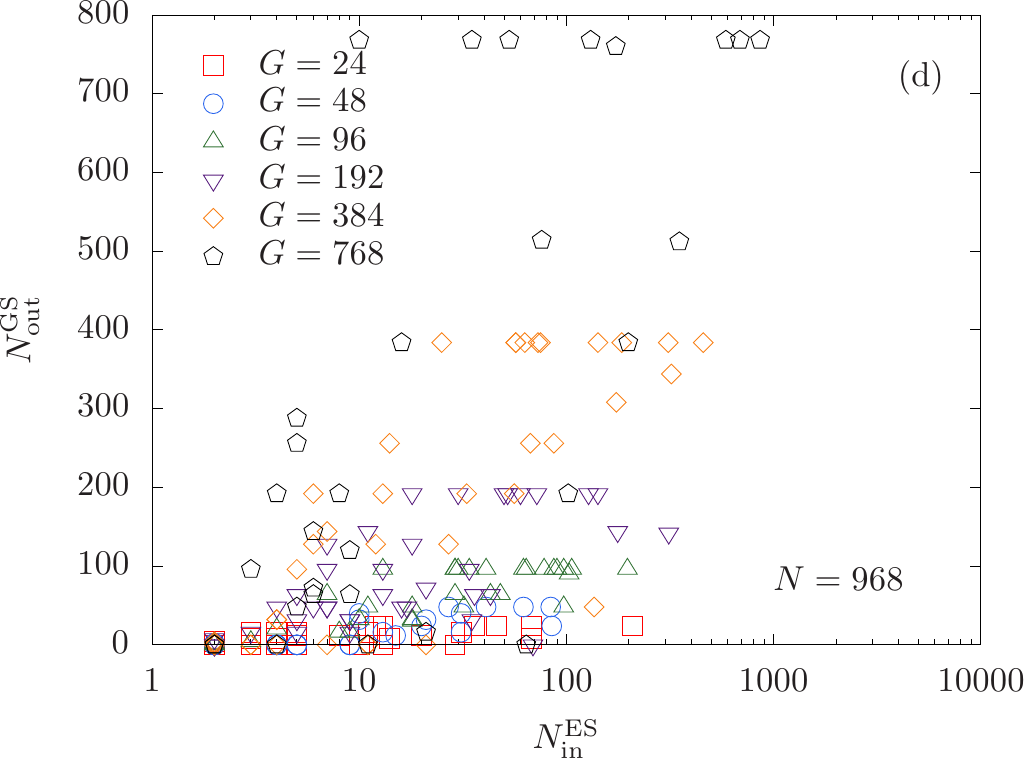}
\caption{
Number of solutions generated using cluster updates
($N^{\rm{GS}}_{\rm{out}}$) as a function of {\em only} excited states found
using the D-Wave 2X quantum annealer ($N^{\rm{ES}}_{\rm{in}}$)
for different system sizes $N$ and known degeneracies $G$: (a) $N = 512$, (b) $N =
648$, (c) $N = 800$, and (d) $N = 968$. Each point
represents an individual instance.  For all points that lie above the
horizontal axis, the resampled excited states produced new minimizing
configurations.
}
\label{fig:C_ES}
\end{figure*}

Figure \ref{fig:C_GS_ES} shows scatter plots of individual instances
produced using the D-Wave 2X quantum annealer for a given degeneracy $G$
and system size $N$. Here $N^{\rm{GS+ES}}_{\rm{in}}$ configurations
(including first-excited states) are fed into the algorithm. The
vertical axis represents the resulting ground-state configurations
$N^{\rm{GS}}_{\rm{out}}$. The horizontal axis only depicts the
ground-state configurations $N^{\rm{GS}}_{\rm{in}}$ fed into the cluster
update for better comparison with the data in Fig.~\ref{fig:C_GS}.
Especially for large $G$ and $N$, the inclusion of excited states
improves the sampling. Figure \ref{fig:C_ES} goes to the extreme by only
allowing nonminimizing first-excited states as input to the algorithm.
As can be seen, the minimizing configurations can often be obtained from
excited states alone. Only in a handful of cases (points on the
horizontal axis) did the cluster updates not generate a minimizing
configuration. This has an important consequence: Intrinsically bad data
from a poor optimization technique can be postprocessed to find a
minimizing configuration of the cost function, provided enough
low-energy input states are available. We accomplish this by keeping
track of the configurations with the lowest energy found.

Next we attempt to systematically study the effects of a poor
initial ground-state pool by restricting the initial set of minimizing
configurations by a fraction $p$. In our experiments we use fractions of
$25\%$ ($p = 0.25$), $50\%$ ($p = 0.50$), and $75\%$ ($p = 0.75$). As
done for the original data set (Table \ref{tab:mandra}), we run the
cluster updates $2^{20}$ times for each data set.

\begin{figure}
\centering
\includegraphics[width=0.45\textwidth]{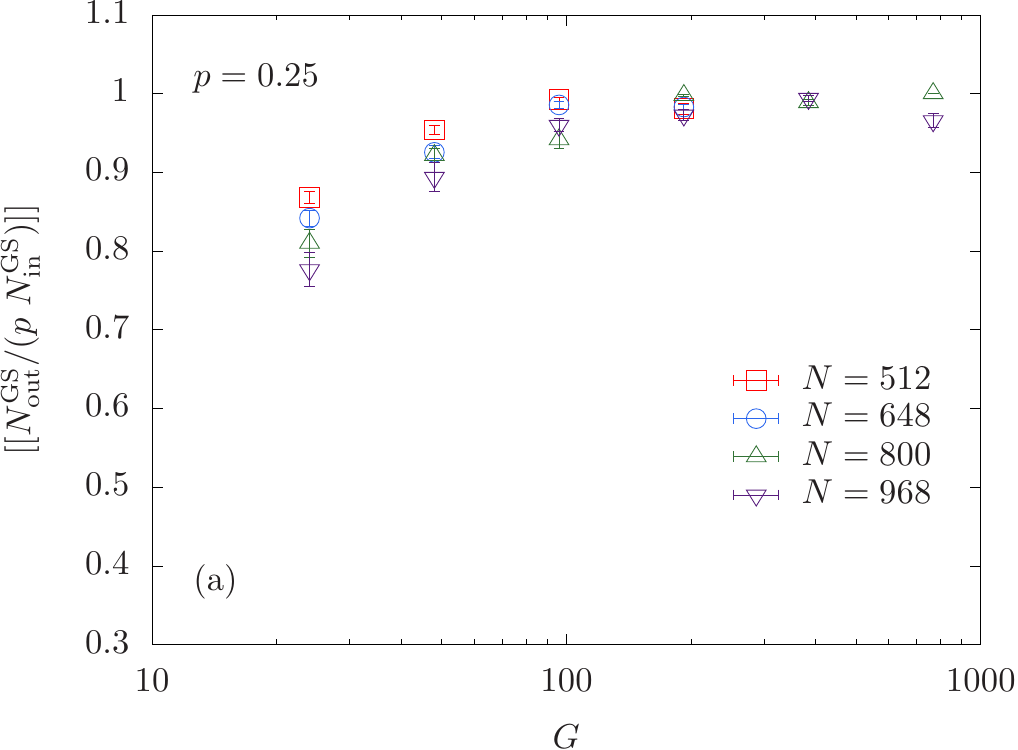}\\[1.5em]
\includegraphics[width=0.45\textwidth]{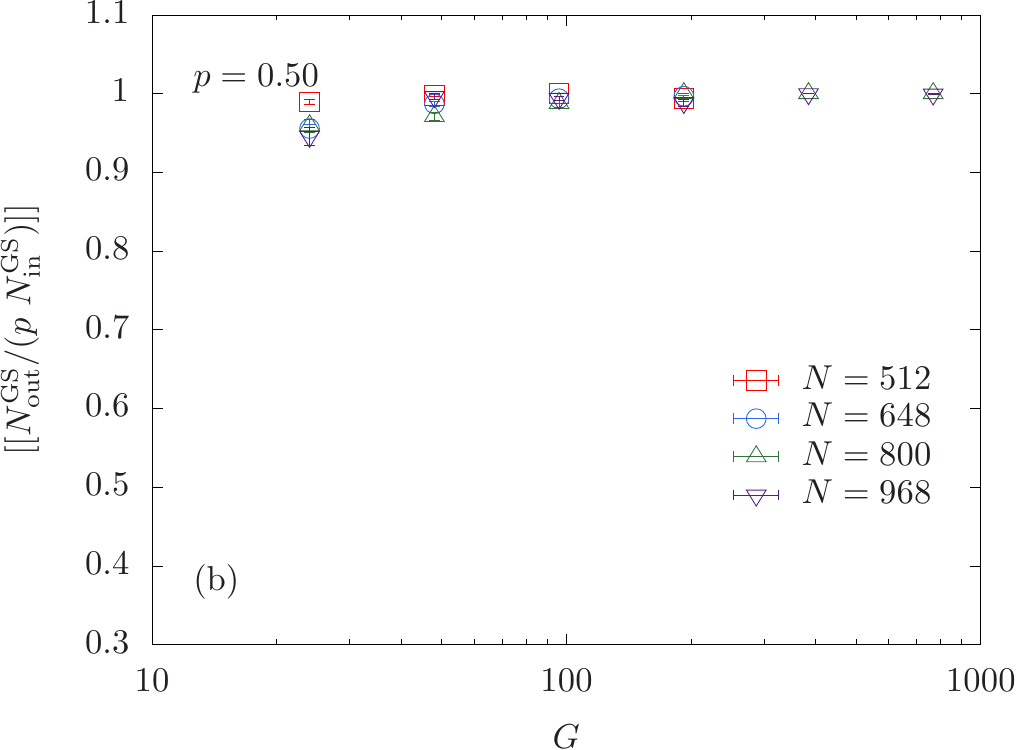}\\[1.5em]
\includegraphics[width=0.45\textwidth]{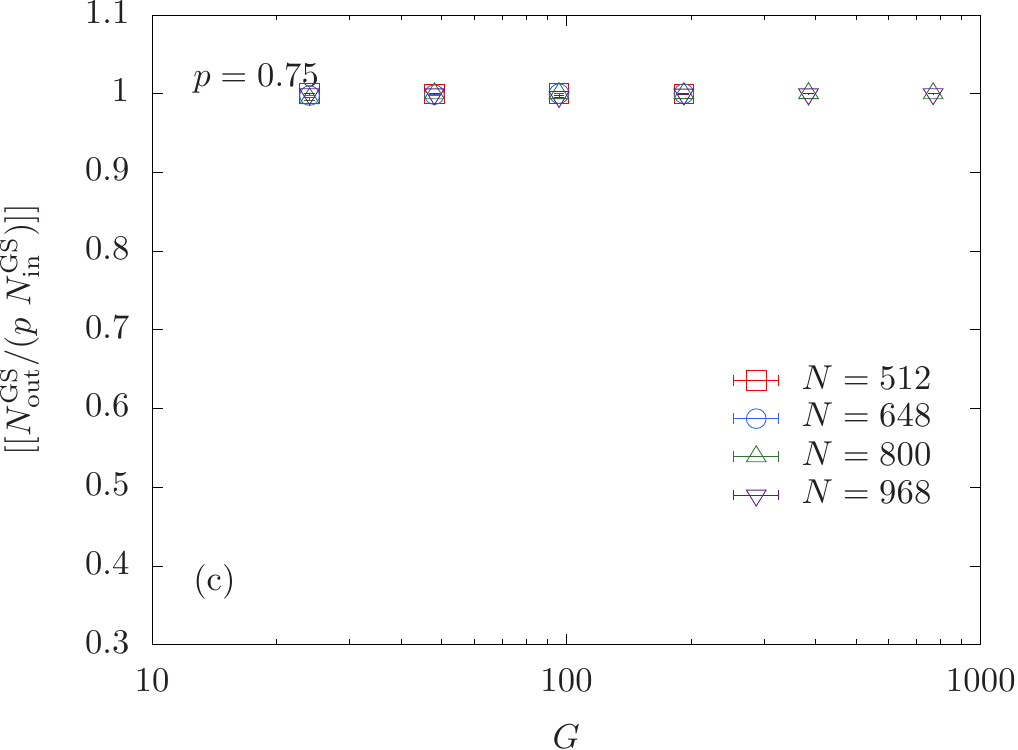}
\caption{ 
Relative improvement $\bm{[}[N^{\rm{GS}}_{\rm{out}} / (p
N^{\rm{GS}}_{\rm{in}})]\bm{]}$ averaged over instances and ten independent
runs as a function of known degeneracy $G$ for different system sizes
$N$ and only using a fraction $p$ of available states after
postprocessing the minimizing configurations obtained with the D-Wave
2X quantum annealer: (a) $p = 0.25$, (b) $p = 0.50$, and (c) $p = 0.75$.  The data are normalized such that for
$\bm{[}[N^{\rm{GS}}_{\rm{out}} / (p N^{\rm{GS}}_{\rm{in}})]\bm{]} = 1$ all known
solutions in the initial ground-state pool are found.  Double-square
brackets $[[\ldots]]$ represent averages over instances and ten
independent random trials.
}
\label{fig:C_P}
\end{figure}

Figure \ref{fig:C_P} shows the relative improvement
$\bm{[}[N^{\rm{GS}}_{\rm{out}} / (p N^{\rm{GS}}_{\rm{in}})]\bm{]}$ averaged over
instances and ten independent subsets for different values of $p$ as a
function of known degeneracy $G$ for different system sizes $N$ after
postprocessing the minimizing configurations obtained with the D-Wave
2X quantum annealer.  Note that the data are normalized by a factor $p$
such that when $\bm{[}[N^{\rm{GS}}_{\rm{out}} / (p N^{\rm{GS}}_{\rm{in}})]\bm{]}
\to 1$ all known configurations are found. As can be seen, even when
only $25\%$ of input states are used, all configurations can be found
for large enough $G$. This shows that the method is asymptotically
robust.

\begin{figure}
\centering
\includegraphics[width=0.44\textwidth]{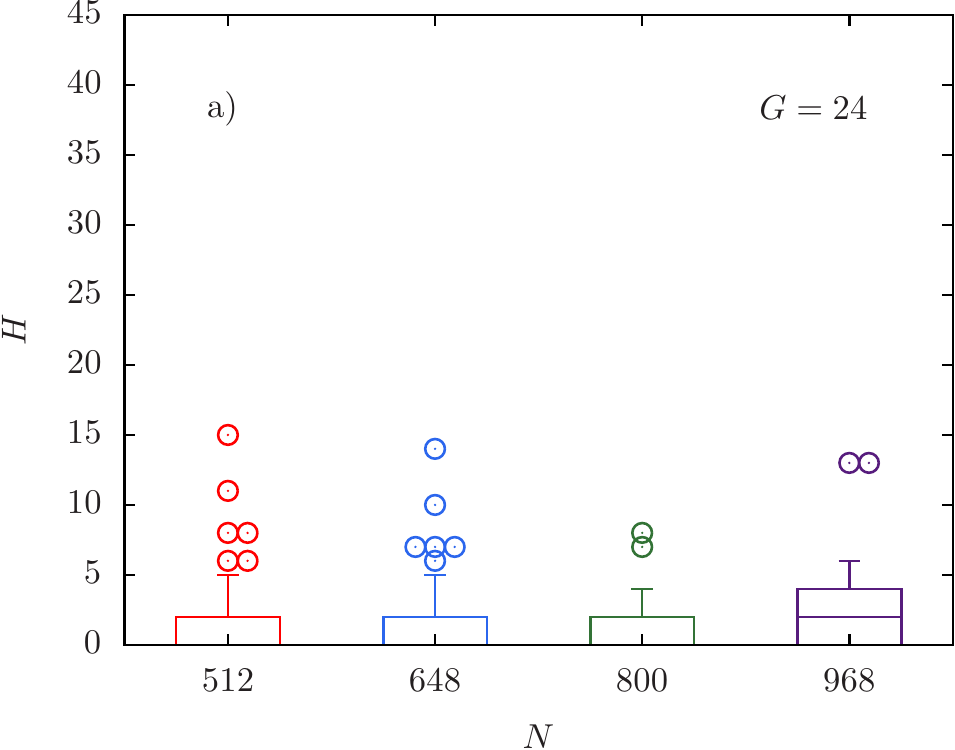}\\[1.5em]
\includegraphics[width=0.44\textwidth]{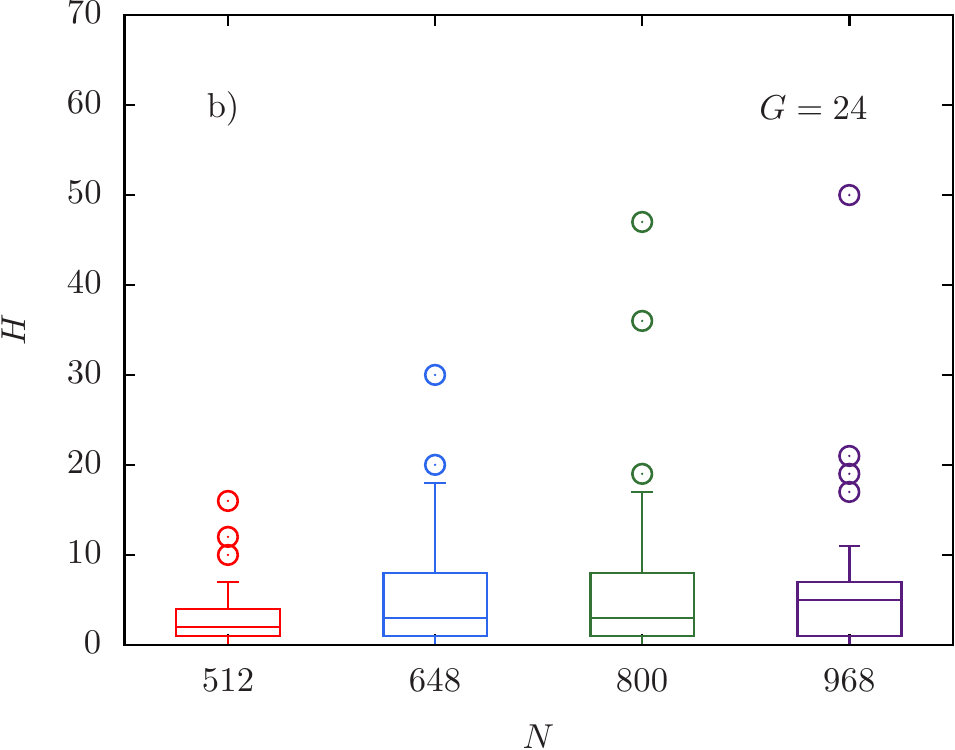}\\[1.5em]
\includegraphics[width=0.44\textwidth]{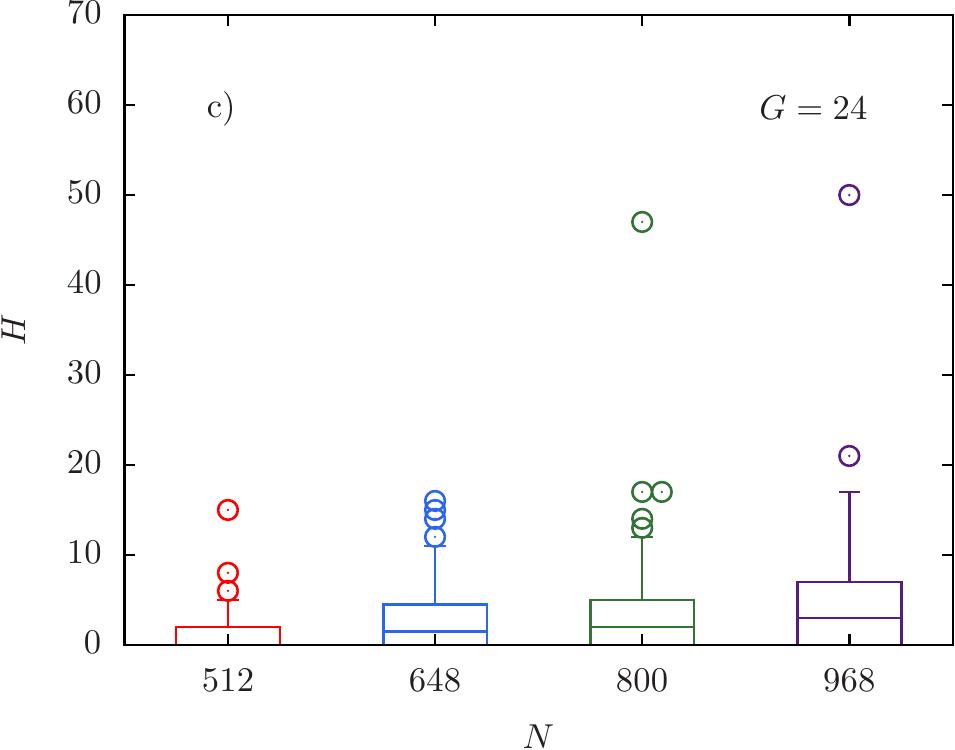}
\caption{
Maximum-minimum hamming distance versus the system size $N$ with degeneracy
$G=24$ for the chimera graph. For each instance the minimum Hamming distance
between each new ground state found and the original pool of configurations is
calculated. The data shown are the maximum of these minima. The box represents
the 25\%-50\% confidence interval, the center line is the median, the bars
are the 5\%-95\% confidence interval, and the points are outliers. Data are shown from initial pools consisting of (a) only ground
states, (b) only excited states, and (c) both ground states and excited states. Outliers represent non-trivial solutions with large Hamming
distance from the initial pool found by the method.
}
\label{fig:hamk2}
\end{figure}

\begin{figure}
\centering
\includegraphics[width=0.44\textwidth]{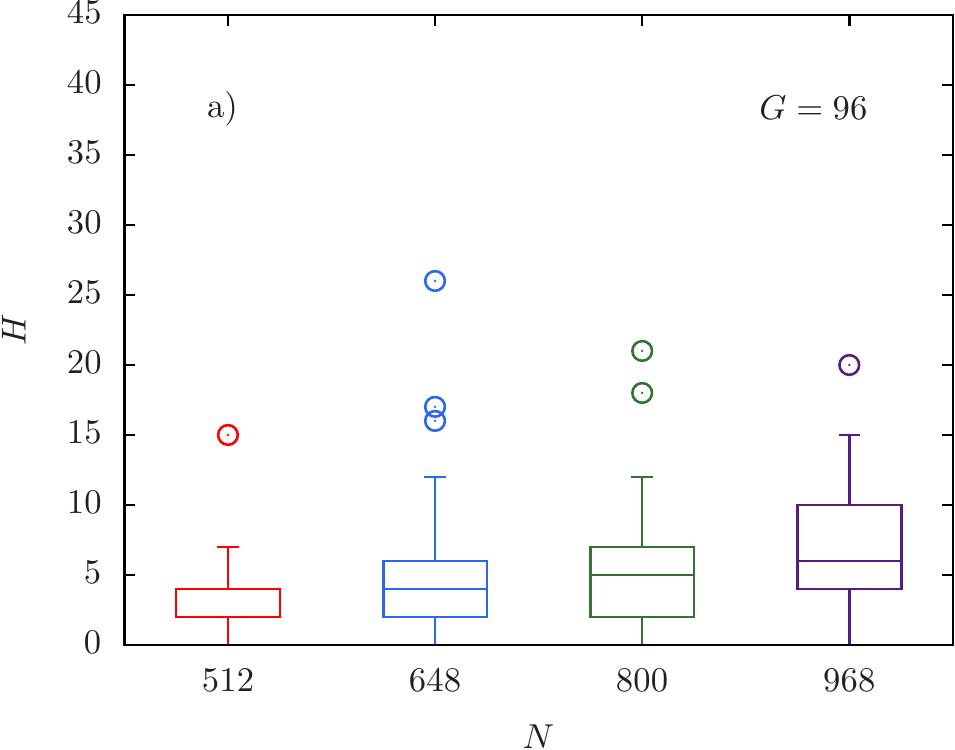}\\[1.5em]
\includegraphics[width=0.44\textwidth]{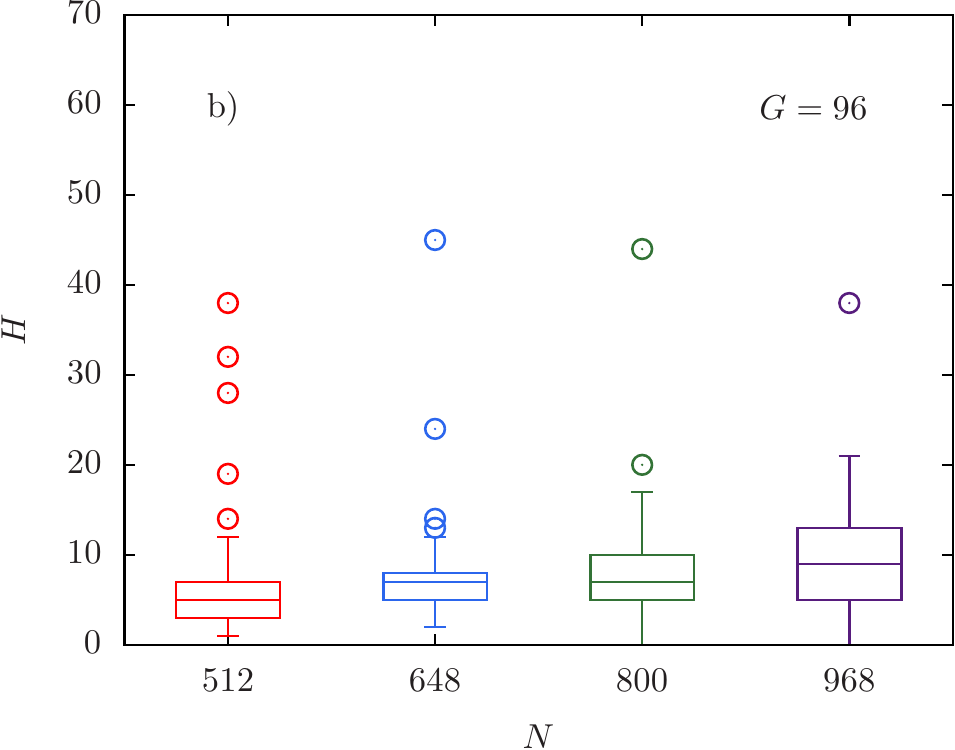}\\[1.5em]
\includegraphics[width=0.44\textwidth]{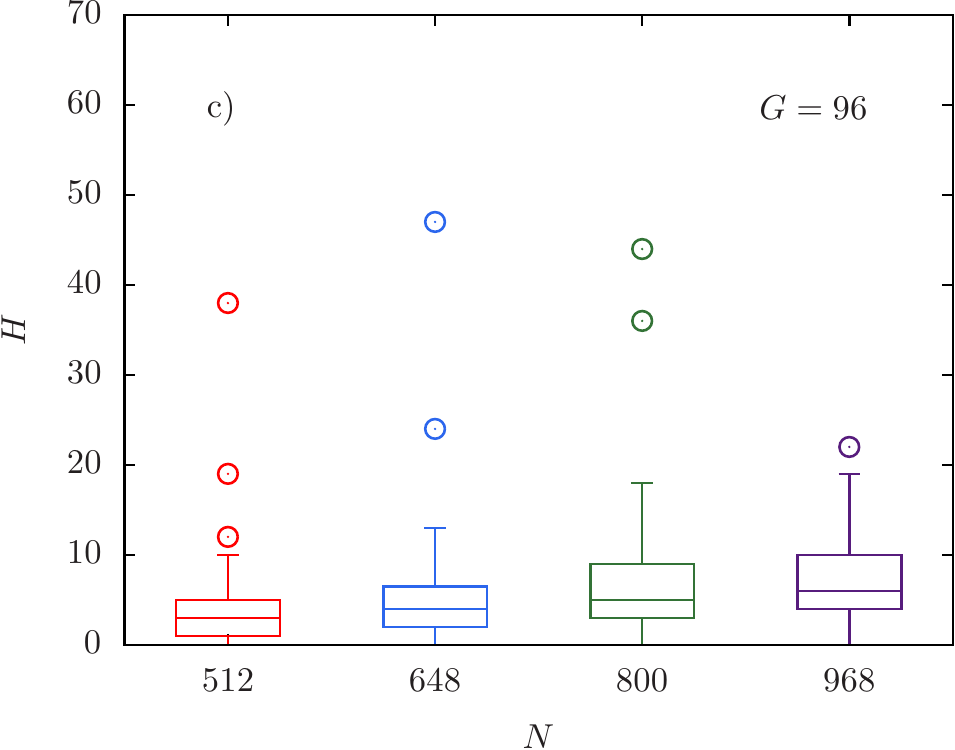}
\caption{ 
Maximum-minimum hamming distance versus the system size $N$ with degeneracy
$G=96$ for the chimera graph. For each instance the minimum Hamming distance
between each new ground state found and the original pool of configurations is
calculated. The data shown are the maximum of these minima. The box represents
the 25\%-50\% confidence interval, the centerline is the median, the bars
are the 5\%-95\% confidence interval, and the points are outliers. Data are shownfrom initial pools consisting of (a) only ground
states, (b) only excited states, and (c) both ground states and excited states. Outliers represent nontrivial solutions with large Hamming
distance from the initial pool found by the method.
}
\label{fig:hamk4}
\end{figure}

Finally, we illustrate the diversity of solutions found by the method. For each
instance, we calculate the minimum Hamming distance of each new ground state
from the pool of original ground states. From these minimum Hamming distances
we take the maximum and analyze the maximum Hamming distance for each instance.
This maximum-minimum Hamming distance $H$ shows if the method can
find solutions which are not close in Hamming distance to the original pool.  

Figures \ref{fig:hamk2} and \ref{fig:hamk4}  show the maximum-minimum Hamming
distance versus system size with degeneracy $G=24$ and $G=96$. The figures
represent whether only ground states [Figs.~\ref{fig:hamk2}(a) and \ref{fig:hamk4}(a)], only excited states [Figs.~\ref{fig:hamk2}(b) and \ref{fig:hamk4}(b)], or both
ground states and excited states [Figs.~\ref{fig:hamk2}(c) and \ref{fig:hamk4}(c)] are included in the initial pool of
states.  The box represent the 25\%-50\% confidence interval, the centerline is the median, the bars are the 5\%-95\% confidence interval, and the
points are outliers. Solutions found by the method are typically close to the
original pool of solutions. In Figs.~\ref{fig:hamk2} and \ref{fig:hamk4} the
median increases with the system size to approximately $H=8$,  which is
consistent with the structure of the chimera Hamiltonian. Increasing the
degeneracy from $G=24$ to $G=96$ also shows an improvement in the median
Hamming distance. Nevertheless, there are many outliers that represent the
diversity of solutions that would likely not be found with low-depth search
starting from the initial pool of configurations, thus demonstrating the
ability of the method to discover nontrivial solutions.

Summarizing, our results show that the resampling can overall improve
ground-state data produced by the D-Wave 2X quantum annealer. However,
these results can be applied more generally to any heuristic.

\begin{table}
\caption{
Number of disorder instances $N_{\rm s}$ sorted by system size $N$ and
number of ground states $G$ in three space dimensions. For each system
size and ground-state degeneracy, the cluster update was applied
$2^{20}$ times to the data set to produce new states for each study
listed below \cite{comment:cpu}.
\label{tab:3d}
}
\begin{tabular*}{\columnwidth}{@{\extracolsep{\fill}} l c c c c c c r}
\hline
\hline
$N$ & $G\!=\!8$ & $G\!=\!12$ & $G\!=\!16$ & $G\!=\!20$ & $G\!=\!24$ &$G\!=\!32$ & $N_{\rm s}$\\
\hline
$64$  & $165$  & $49$ & $13$ & $8$  & $3$  & $4$  & $252$\\
$125$ & $313$ & $105$ & $89$ & $22$ & $37$ & $15$  & $581$\\
$216$ & $376$ & $139$ & $198$ & $23$ & $107$ & $70$ & $913$\\
\hline
\hline
\end{tabular*}
\end{table}

\subsection{Three-dimensional lattices}
\label{sec:3d}

To study the effects of dimensionality, we now show experiments for
three-dimensional cubic lattices. Our motivation lies in the fact that,
in general, cluster updates become inefficient for dense graphs
\cite{zhu:15b,houdayer:01}. However, due to the intrinsic frustration of
the problems, as well as the fact that we apply the updates at either 
zero or close-to-zero temperature, the cluster updates efficiently
produce new states. Instances for $N = 64$, $125$, and $216$ sites and
couplers drawn from $J_{ij} \in \{\pm 5, \pm 6, \pm 7\}$ are initially
optimized exactly using the spin-glass server \cite{juenger:sg}.  Note,
however, that the spin-glass server only gives one minimizing
configuration.  To produce the data sets for resampling, we use
simulated annealing \cite{kirkpatrick:83} (Isakov {\em{et al.}} implementation
\cite{isakov:15} with $T_{\rm max} = 10$, $T_{\rm min} = 0.33$, $2^{14}
$ sweeps, and $10^5$ repetitions for each sample). Instance parameters
are listed in Table \ref{tab:3d}.  Because the problems are relatively
small, simulated annealing (even with a poor choice of parameters) tends
to find all minimizing configurations.  As such, we generated a
synthetic data set where we only used first excited-states to verify
that the cluster updates can use this information to find minimizing
configurations.

\begin{figure}
\centering
\includegraphics[width=0.45\textwidth]{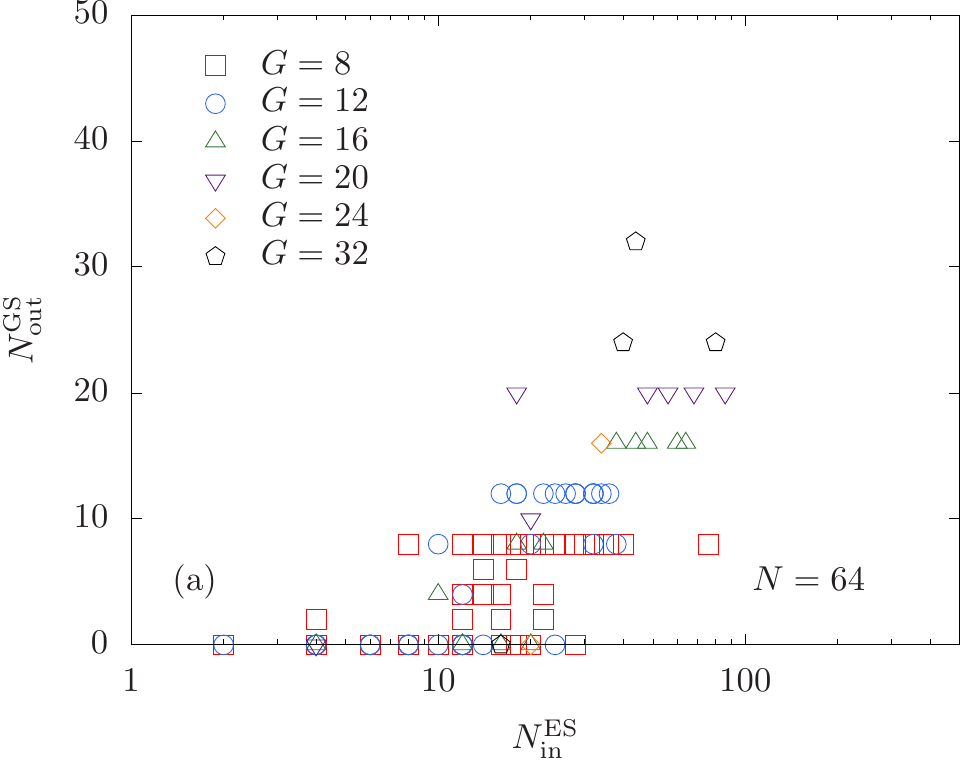}\\[0.75em]
\includegraphics[width=0.45\textwidth]{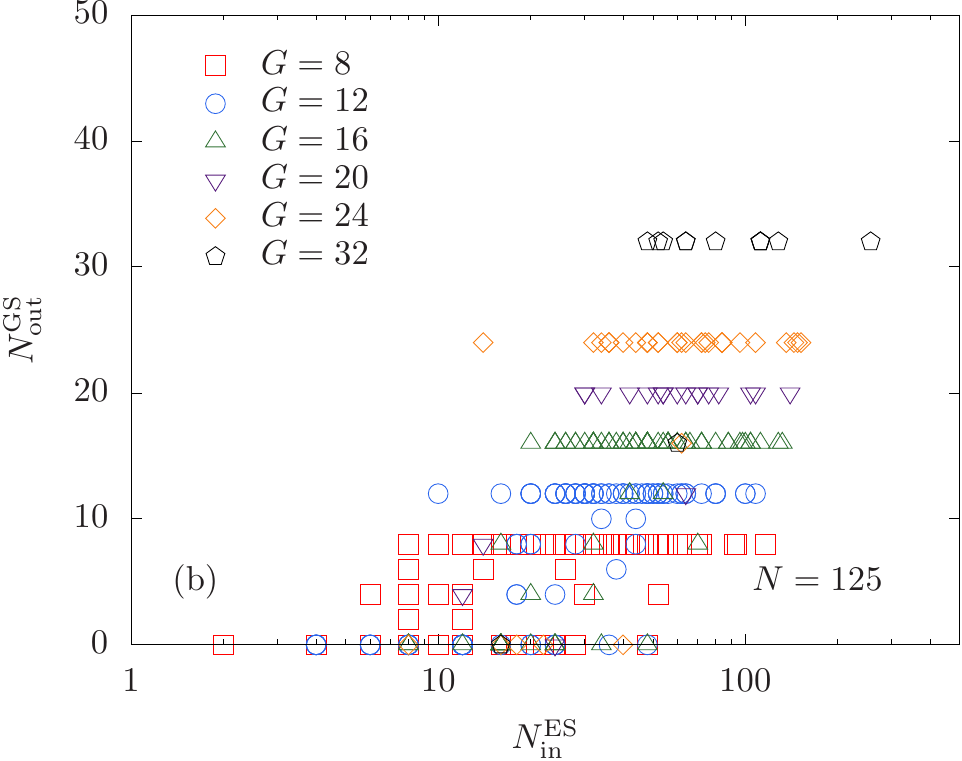}\\[0.75em]
\includegraphics[width=0.45\textwidth]{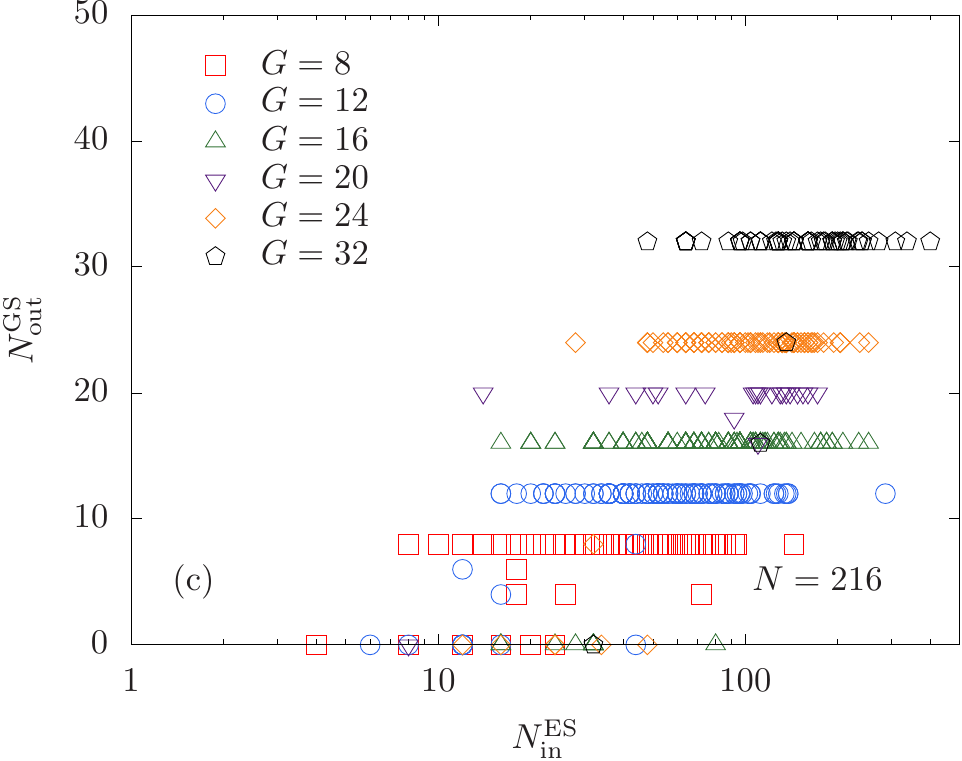}
\caption{
Number of solutions generated using cluster updates
($N^{\rm{GS}}_{\rm{out}}$) as a function of only excited states found
using simulated annealing ($N^{\rm{ES}}_{\rm{in}}$)
for different system sizes $N$ and known degeneracies $G$ for three-dimensional systems:  (a)
$N = 64$ ($L = 4$), (b) $N = 125$ ($L = 5$), and (c) $N = 216$ ($L = 6$).  Each point
represents an individual instance.  For all points that lie above the
horizontal axis, the resampled excited states produced new minimizing
configurations. Note that for $N = 64$ most instanced were solved, i.e.,
the pool of excited states is small.
}
\label{fig:3D_GS_ES}
\end{figure}

Figure \ref{fig:3D_GS_ES} shows results using only first-excited states
as the input to the cluster updates in three space dimensions.  As can
be seen, minimizing configurations can be obtained from excited states
only. This demonstrates that the cluster resampling works in space
dimensions where, from a purely geometrical point of view, clusters
would percolate and therefore be inefficient.

\begin{figure}
\centering
\includegraphics[width=0.45\textwidth]{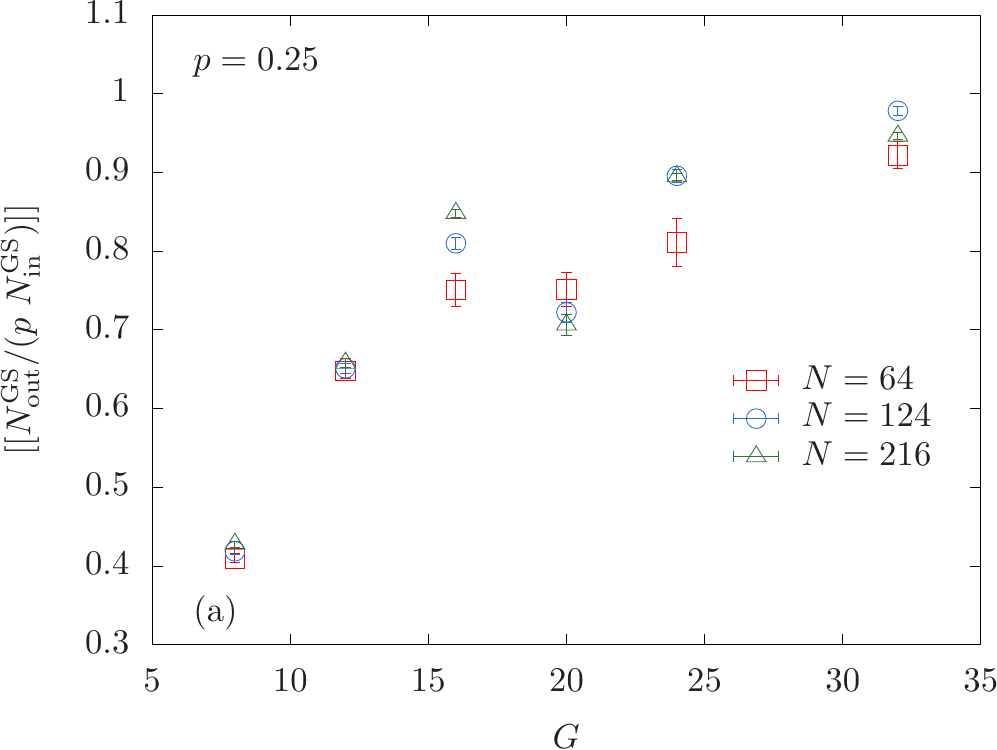}\\[1.5em]
\includegraphics[width=0.45\textwidth]{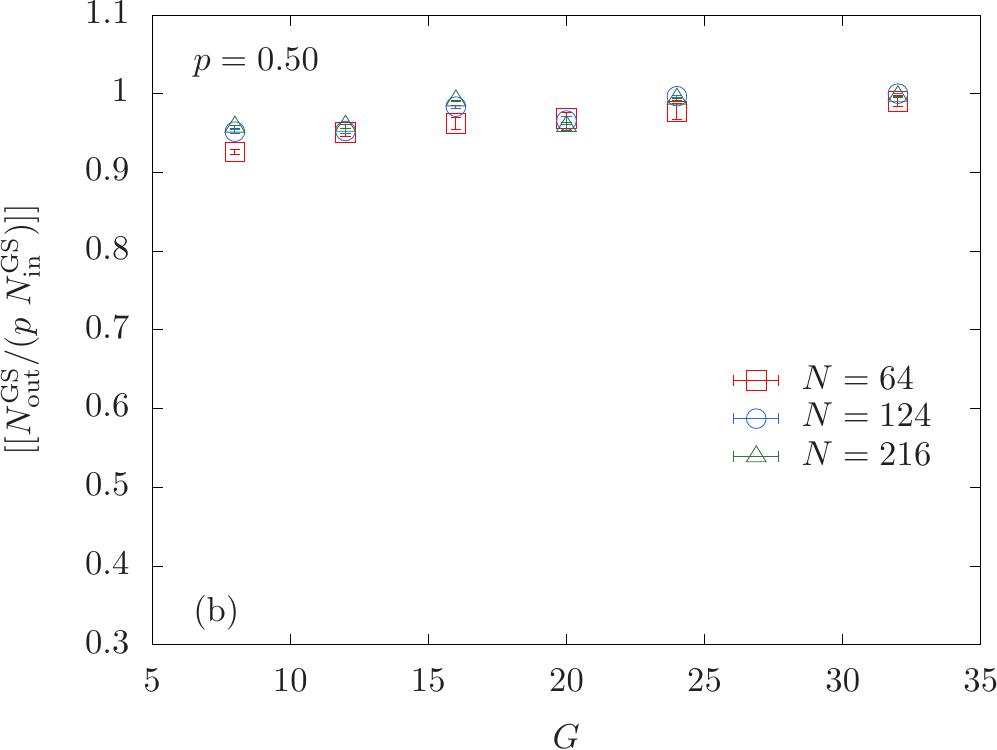}\\[1.5em]
\includegraphics[width=0.45\textwidth]{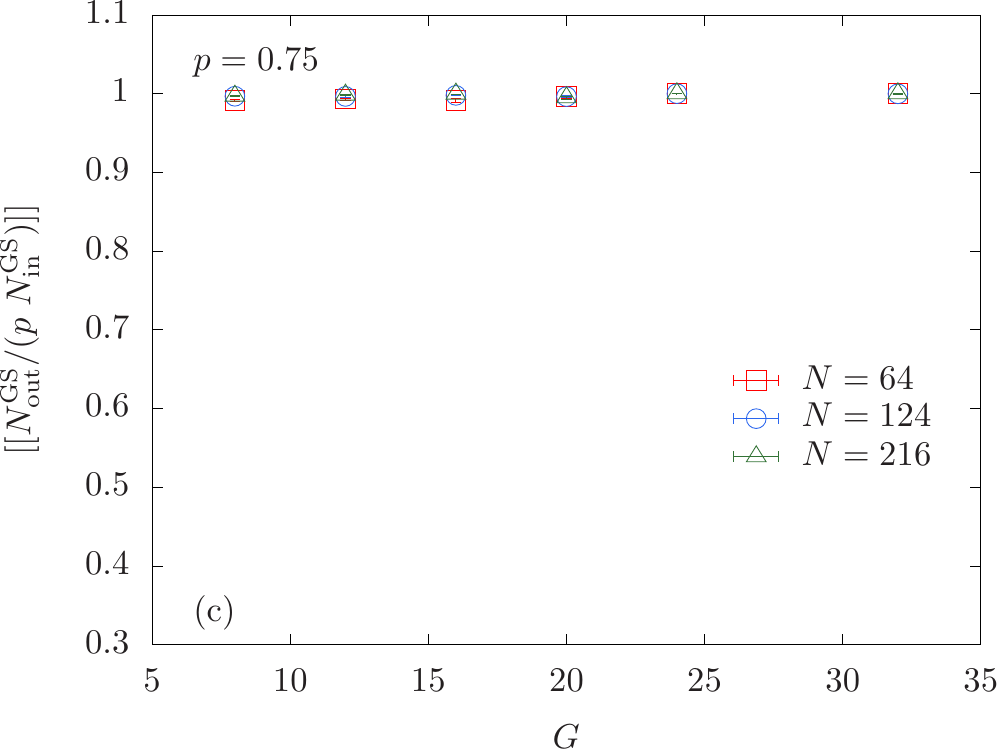}
\caption{ 
Relative improvement $\bm{[}[N^{\rm{GS}}_{\rm{out}} / (p
N^{\rm{GS}}_{\rm{in}})]\bm{]}$ averaged over instances and ten independent
runs as a function of known degeneracy $G$ for different system sizes
$N$ and only using a fraction $p$ of available states after
post-processing the minimizing configurations obtained with simulated annealing
in three space dimensions. (a) $p = 0.25$, (b) $p = 0.50$, and
(c) $p = 0.75$. The data are normalized such that for
$\bm{[}[N^{\rm{GS}}_{\rm{out}} / (p N^{\rm{GS}}_{\rm{in}})]\bm{]} = 1$ all known
solutions in the initial ground-state pool are found.  Double-square
brackets represent averages over instances and ten
independent random trials.
}
\label{fig:3D_P}
\end{figure}

As done for the chimera topology in Sec.~\ref{sec:chimera}, we use
fractions of $25\%$ ($p = 0.25$), $50\%$ ($p = 0.50$), and $75\%$ ($p =
0.75$) of the actual ground-state pool computed with simulated
annealing.  Figure \ref{fig:3D_P} shows the relative improvement
$\bm{[}[N^{\rm{GS}}_{\rm{out}} / (p N^{\rm{GS}}_{\rm{in}})]\bm{]}$ averaged over
instances and ten independent subsets for different values of $p$ as a
function of known degeneracy $G$ for different system sizes $N$ after
postprocessing the minimizing configurations.  Again, the data are
normalized by a factor $p$ such that when $\bm{[}[N^{\rm{GS}}_{\rm{out}} / (p
N^{\rm{GS}}_{\rm{in}})]\bm{]} \to 1$ all known configurations are found. As
can be seen, even when only $25\%$ of input states are used, all
configurations can be found for large enough $G$.

\subsection{Nondegenerate problems}
\label{sec:nondeg}

\begin{figure}
\centering
\includegraphics[width=0.39\textwidth]{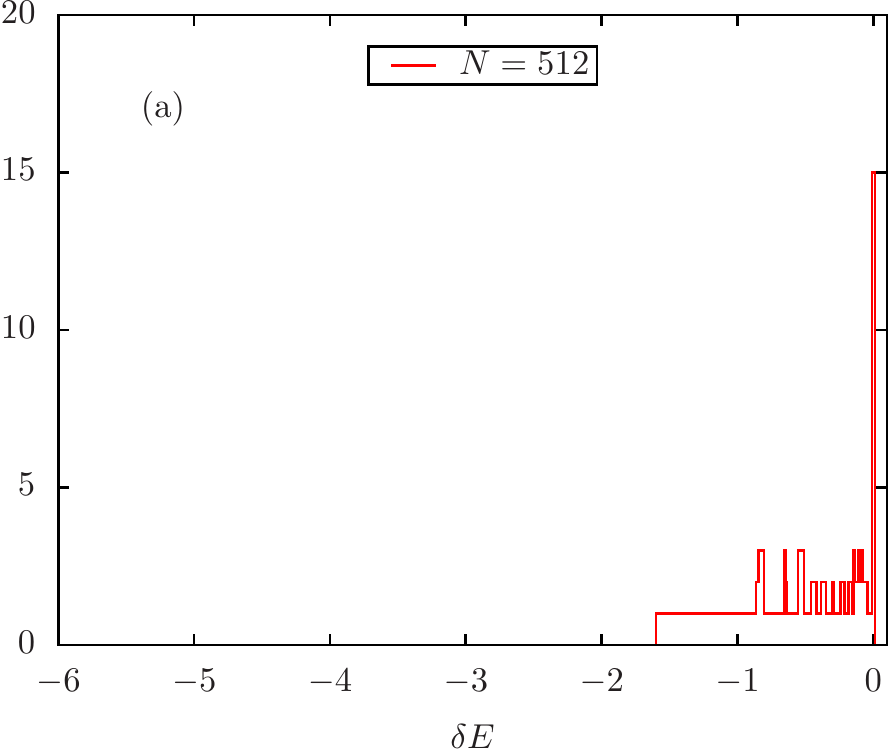}\\[1.5em]
\includegraphics[width=0.39\textwidth]{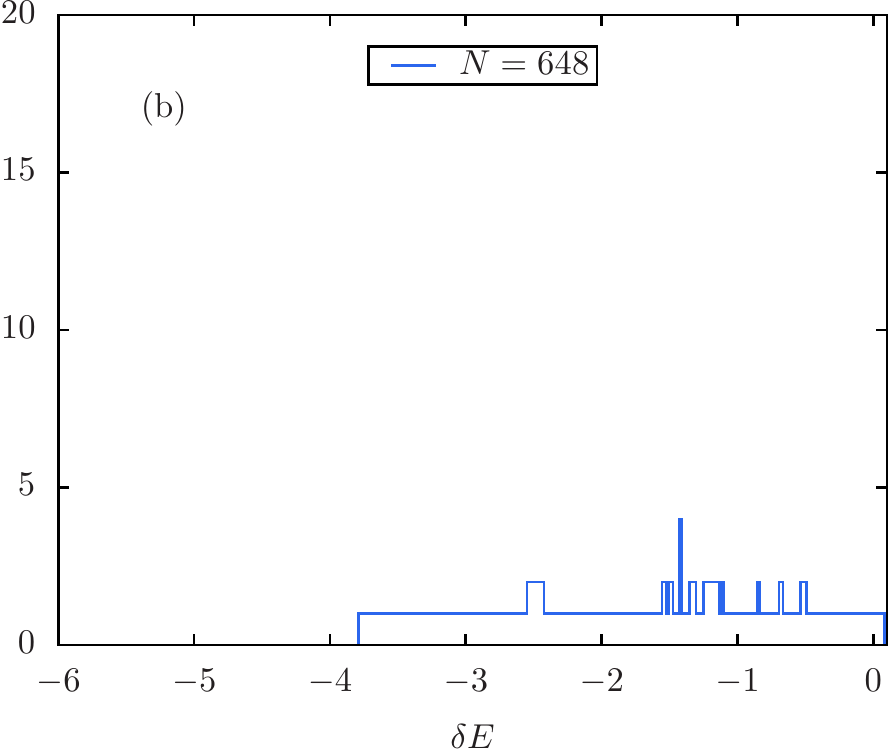}\\[1.5em]
\includegraphics[width=0.39\textwidth]{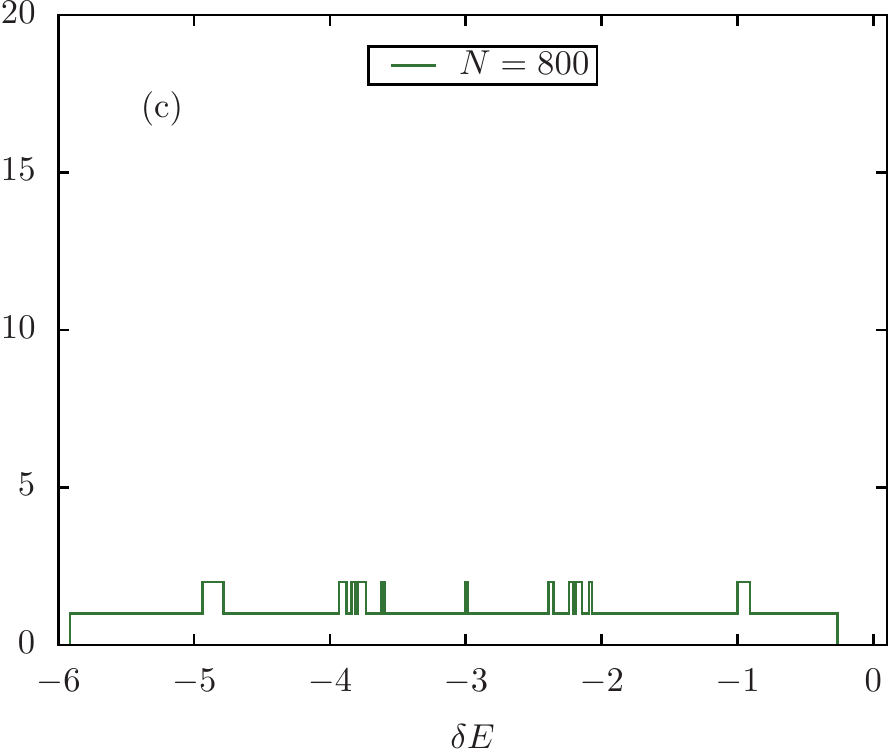}
\caption{
Histograms of the change in energy $\delta E$ between states found by
the D-Wave 2X quantum annealer and the same states post processing them
with the cluster updates for different system sizes $N = 8 \times c^2$
for problems with Gaussian-distributed couplers. (a) $N = 512$, (b) $N
= 648$, and (c) $N = 800$. In most cases
lower-energy configurations were obtained using the cluster algorithm
with the improvement becoming more pronounced for the larger system
sizes.
}
\label{fig:C_G_dE}
\end{figure}

Systems that have a unique ground state tend to be computationally more
difficult than highly degenerate systems \cite{katzgraber:15}. In this
section we study spin-glass Hamiltonians on the chimera lattice with
couplers drawn from a Gaussian distribution with zero mean and unit
standard deviation. These, by construction, have unique ground states.
We generate a data set for different chimera lattice sizes with $N = 8
\times c^2$ variables using the D-Wave 2X quantum annealer. Because
Gaussian couplers require high precision, the D-Wave analog annealer is
notoriously bad at minimizing Gaussian problems.  To illustrate how the
cluster updates can improve low-quality data, in Fig.~\ref{fig:C_G_dE}
we show histograms of the change in energy between the configurations
produced by the D-Wave device and the same configurations after post
processing them with the cluster updates. For each instance, $2^{17}$
resampling steps were performed of the approximately $100$ instances. As
can be seen, the algorithm improves the data considerably and is (in
some cases) able to find the actual ground-state configuration of the
nondegenerate problem.

We have also attempted to find the solutions to these instances using
simulated annealing \cite{isakov:15}. Assuming that the obtained states
are the true minimizing configurations, we study the fraction of solved
problems as a function of system size for the D-Wave data, as well as
the postprocessed data using the cluster algorithm. For example, for $N
= 483$ the fraction of problems solved increased from $0.58$ to $0.73$.
Similarly, for $N = 615$, the fraction of solved problems increased from
$0.11$ to $0.29$. For smaller system sizes no noticeable improvement was
observed, whereas for the largest problems with $758$ variables the
cluster update found lower-energy configurations than simulated
annealing was able to find.

\subsection{Sampling at finite temperature}
\label{sec:finite}

We now study the sampling of states at a finite temperature using the
cluster updates. This is of importance for applications such as machine
learning where a diverse pool of states is needed for the training step.
For two replicas $(1)$ and $(2)$ needed for a cluster update the total energy $\Delta E^{(1)} + \Delta E^{(2)} = 0$ has to be zero.
However, the individual changes $\Delta E^{(1)} = - \Delta E^{(2)}$ can
be nonzero. If these changes are typically small, then the cluster
updates can be used to find uncorrelated samples at (best case) constant
temperature.

\begin{figure} \centering
\includegraphics[width=0.44\textwidth]{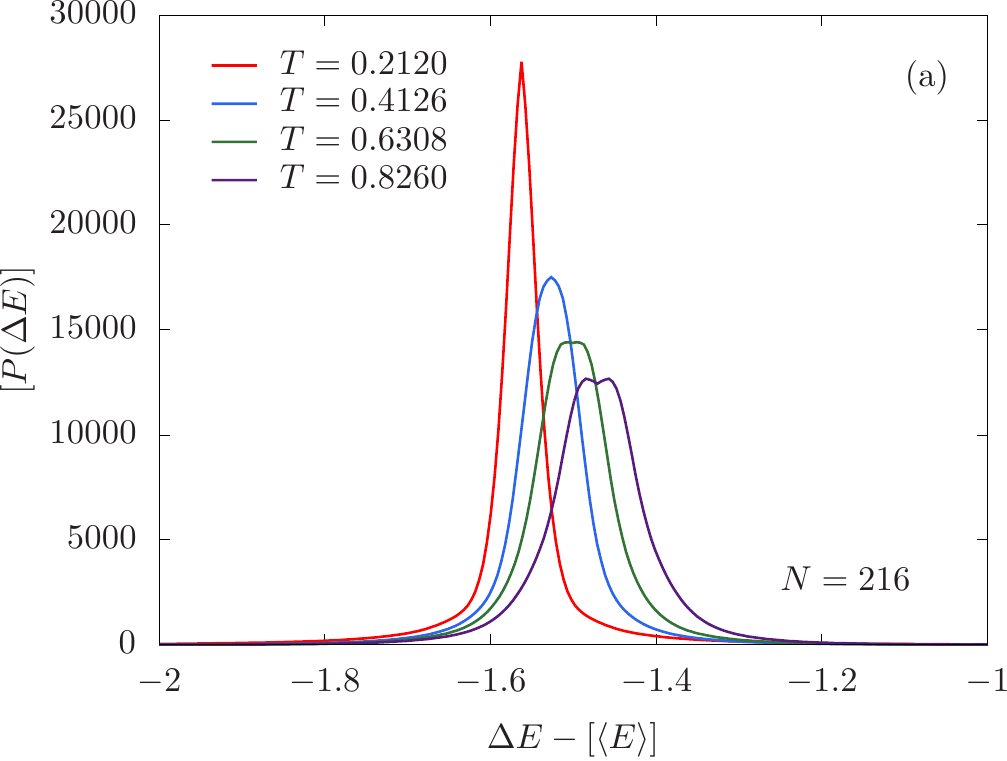}\\[1.5em]
\includegraphics[width=0.45\textwidth]{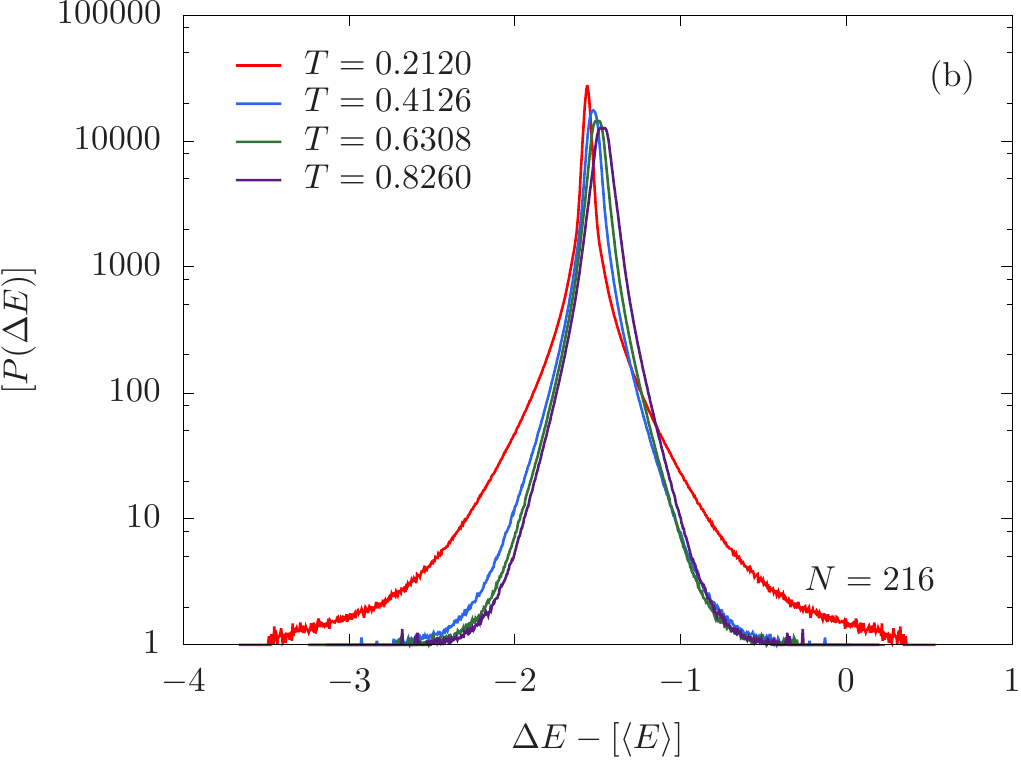} \caption{ Histogram of the
distribution of $\Delta E = |\Delta E^{(1)}| = |\Delta E^{(2)}|$ of cluster
moves around the average thermodynamic energy $\langle E \rangle$ at different
temperatures $T$ in three space dimensions ($N = 216$) during a
finite-temperature Monte Carlo simulation that includes cluster updates. In
general, the changes in energy are small compared to the average energy with
rare large rearrangements. Both panels show the same data set with (b) zooming into the tails.  } \label{fig:3D_dE}
\end{figure}

For this study, we perform a full Monte Carlo simulation of a
three-dimensional Ising spin glass using parallel tempering Monte Carlo
\cite{hukushima:96} as well as isoenergetic cluster updates \cite{zhu:15b}.
System sizes of $N = 4^3 = 64$, $5^3 = 125$, and $6^3 = 216$ are thermalized
using $2^{18}$ Monte Carlo sweeps. For these small system sizes, the instances
are thought to be equilibrated. Measurements are performed over an additional
$2^{16}$ Monte Carlo sweeps.  For the parallel tempering updates, $30$
temperatures in the range $[0.212,1.636]$ are used. However, in this case we
are not interested in the thermal average of observables, except for the energy
per spin $[\langle E \rangle]$ averaged both over disorder and Monte Carlo
time.  During the simulation, we keep track of the change in energy $\Delta E$
for each replica produced by a cluster update for a given sample and bin the
data. Figure \ref{fig:3D_dE} shows histograms for $L=6$ of $\Delta E = |\Delta
E^{(1)}| = |\Delta E^{(2)}|$ averaged over disorder. The horizontal axis has
been shifted by the average energy per spin to better highlight the relative
magnitude of the change. Both panels in the figure show the same data, with
Fig.~\ref{fig:3D_dE}(b) using a vertical logarithmic scale to highlight the
tails of the distribution. As can be seen, while there are pronounced tails,
which corresponds to large changes in the energy of a given replica, the vast
majority of changes in the energy are comparably small after a cluster update.
Combining the cluster updates with a simple postprocessing where samples are
only stored that have $\Delta E$ within a desired window, results in an
efficient sampling of states at almost constant finite temperature.

\subsection{Improvement of fair sampling on D-Wave quantum annealers}
\label{sec:sampling}

Reference \cite{mandra:17} demonstrated that transverse-field quantum
annealing on the D-Wave 2X quantum annealer is not a fair sampling 
heuristic.
In fact, because the problems studied have couplers in the range
$J_{ij} \in \{\pm 5, \pm 6, \pm 7\}$, the energy gap between states
is $\Delta E = 2/7$. Thus one can perform detailed fair sampling
statistics not only of the ground states, but also for excited states.

In order to better appreciate the exponential suppression of sampling on
the D-Wave quantum device, it is possible to introduce the observable
$\Theta_\text{max}$ defined as the maximum absolute difference of the
empiric cumulative distribution $\tilde F(x)$ with respect to the
cumulative of a uniform distribution $F(x)$, namely, $\Theta_\text{max}
= \max_{x} |\tilde F(x) - F(x)|$, with $x$ the state index
\cite{mandra:17}. The test (which is similar in purpose to the
Kolmogorov-Smirnov test) is useful to understand how close an empirical
distribution is to the expected distribution. More precisely, the
smaller $\Theta_\text{max}$ is, the more similar the distributions are.
In general, the number of states at fixed energy $E$ that the D-Wave
quantum annealer can find widely varies from instance to instance.  To
overcome this limitation, it is possible to compare the empiric
$\Theta_\text{max}$ with a ``baseline'' computed by using an amount of
uniformly distributed random numbers which is equal to the number of
states with a given energy $E$ one was able to find \cite{mandra:17}. This
baseline plays an important role since it allows one to take into
account finite-size effects that would erroneously give a false bias.

\begin{figure*}
\centering
\includegraphics[width=0.45\textwidth]{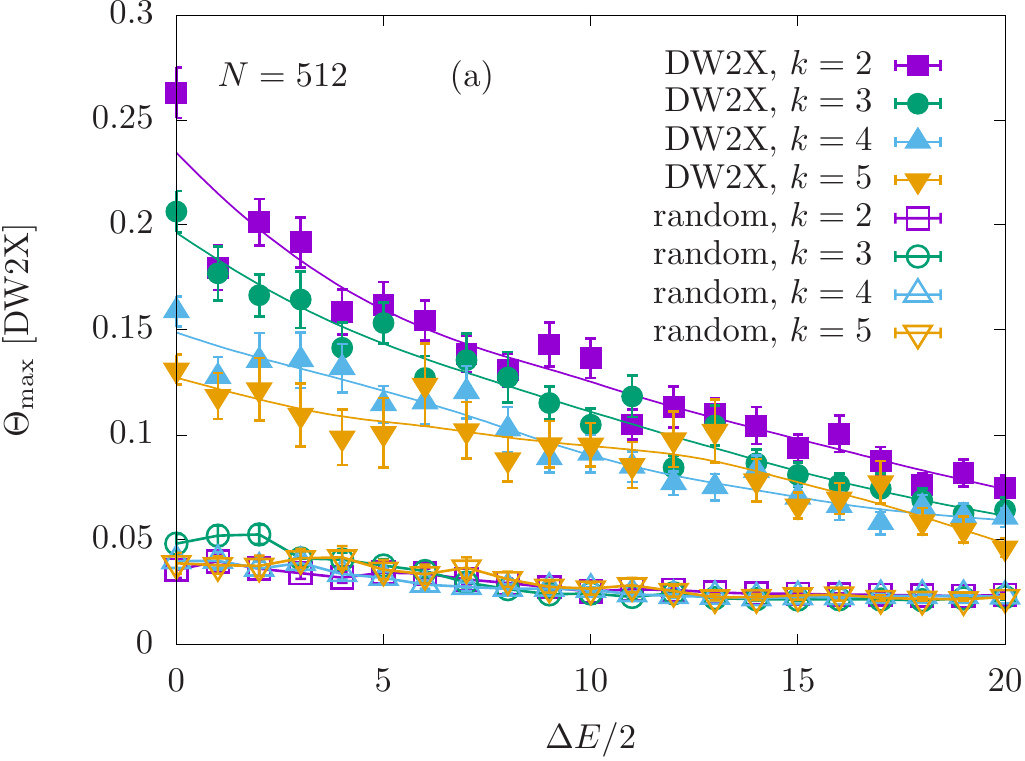}
\includegraphics[width=0.45\textwidth]{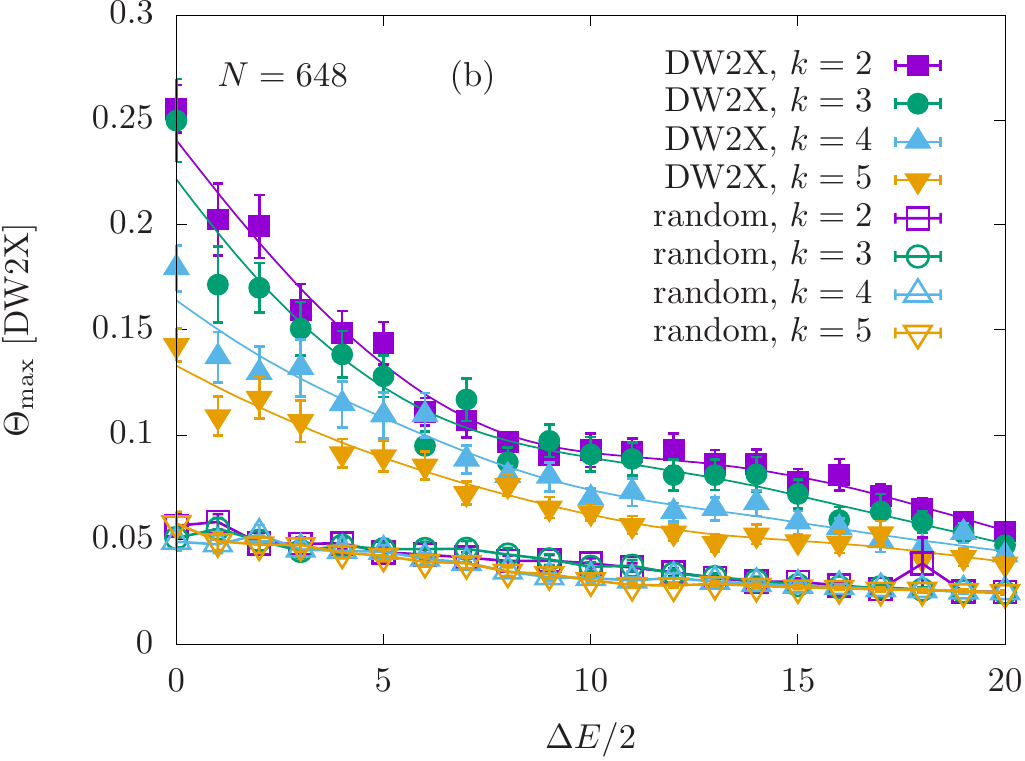}
\caption{
Comparison of $\Theta_\text{max}$ (maximum absolute difference of the
empirical cumulative distribution with respect to the expected
distribution, see the text) for the distribution of states found by the
D-Wave 2X (DW2X) device at a given energy $E = E_\text{gs} + \Delta E$ for $N=512$ and $N=648$.
The baseline for $\Theta_\text{max}$ is computed by using a random
uniform distribution. As one can see, the bias in sampling for the
D-Wave 2X device persists up to the $20^\text{th}$ level when
$\Theta_\text{max}$ for the quantum annealer is comparable to the
baseline.
}\label{fig:theta_ex_states}
\end{figure*}

Figure~\ref{fig:theta_ex_states} shows the comparison of
$\Theta_\text{max}$ for the distribution of states found by the D-Wave
2X device at a given energy $E = E_\text{gs} + \Delta E$, where
$E_\text{gs}$ is the ground-state energy and $\Delta E$ are discrete
(integer) energy shifts in multiples of $2/7$ above the ground state.
As one can see, the bias not only is limited to the ground state, as
reported in Ref.~\cite{mandra:17}, but is persistent up to the
$20^\text{th}$ excited state.

\begin{figure*}
\centering
\includegraphics[width=0.45\textwidth]{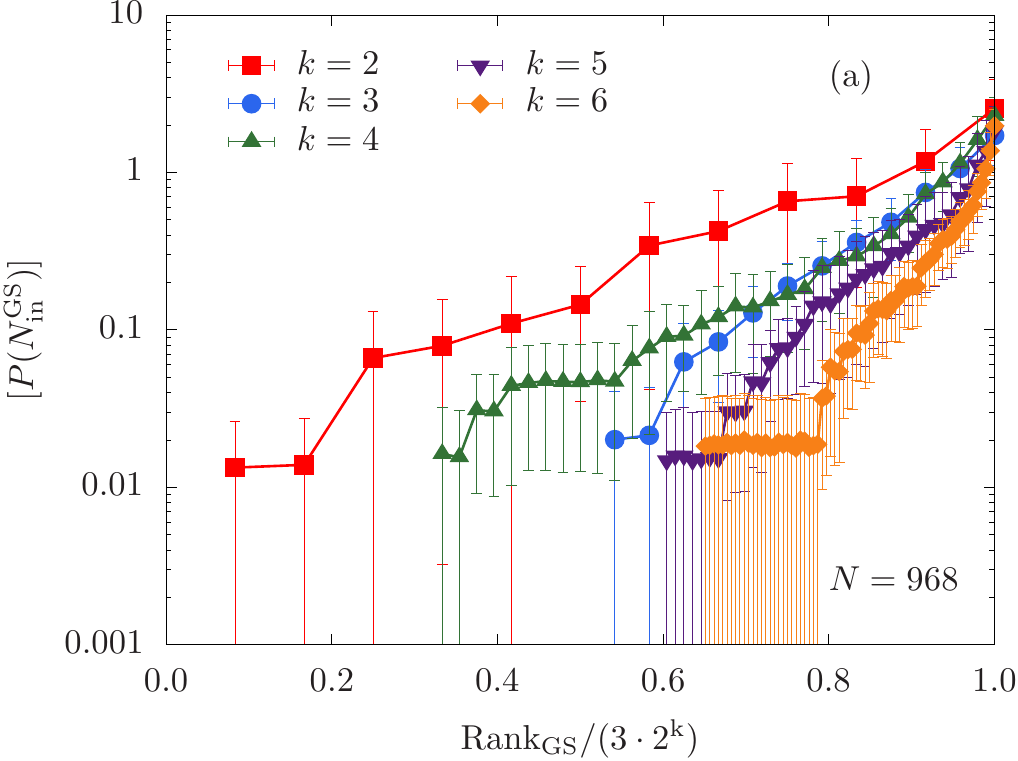}
\includegraphics[width=0.45\textwidth]{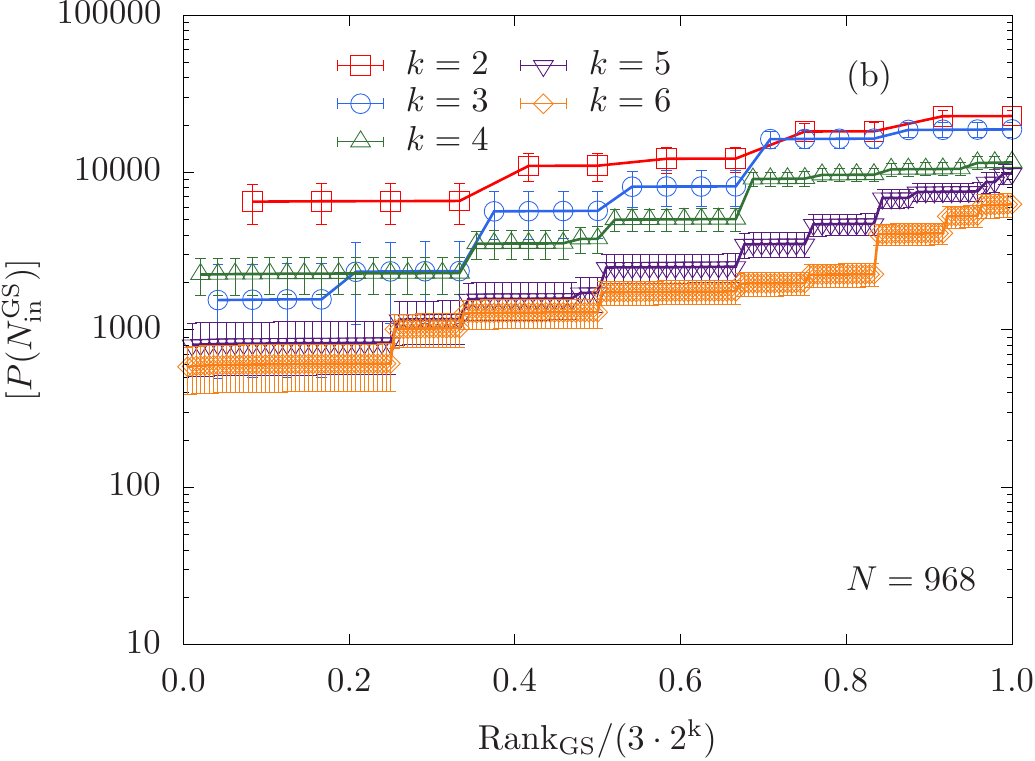}
\caption{
(a) Binned number of times a ground-state configuration is found using the
D-Wave 2X quantum annealer sorted by rank compared to (b) the
cluster-update post-processed data.  Data for instances on
a chimera lattice with $N$ sites and $G = 3\times2^k$ ground states are
shown. The horizontal axis is normalized by $G$ for easier display of
the data. While the magnitudes of the vertical axes are different in
both panels, in both cases the range spans four orders of magnitudes.
This allows for a direct visual comparison of the slopes. The raw data
show a clear bias in (a) that is reduced considerably after
applying the cluster updates shown in (b).
}\label{fig:bias}
\end{figure*}

We reanalyzed the data of Ref.~\cite{mandra:17} using $2^{17}$ cluster
updates. Figure \ref{fig:bias} shows a side-by-side comparison of the data
computed on the D-Wave device in Fig.~\ref{fig:bias}(a) to the postprocessed
data using the cluster updates in Fig.~\ref{fig:bias}(b). Problems have been
selected to have $G = 3 \times 2^k$ ground-state degeneracy. As can be seen
clearly, the postprocessed data show less bias, i.e., the cluster updates can
be applied to biased data sets to reduce biased sampling. By increasing the
number of samples, the limiting distribution of uniform sampling follows
${\rm{erf}}^{-1}$ as shown in the Appendix.

\section{Summary and outlook}
\label{sec:conc}

We have presented a cluster update routine that can vastly improve data
sampling in polynomial time. The approach is based on isoenergetic
cluster updates \cite{zhu:15b}, first introduced for two-dimensional
lattices by Houdayer \cite{houdayer:01}. Using experimental data
produced with the D-Wave 2X quantum annealer as well as synthetic data
on a three-dimensional cubic lattice, we demonstrated different
approaches to apply the cluster updates to improve sampling at both 
zero and finite temperatures. In particular, we demonstrated how
the ground-state manifold of degenerate problems can be sampled more
efficiently, as well as how finite-temperature data for, e.g., machine
learning applications, can be used to produce more samples at either the
same or similar temperature. We emphasize that the approach is generic
and thus can be extended to many problems across disciplines. Application of resampling to machine learning and related applications is left to future work.

We recently became aware of the work in
Ref.~\cite{dorband:18x}. The approach is similar
in nature to the work proposed here (also described in
Refs.~\cite{ochoa:17,katzgraber:18x}), however it only
focuses on reducing the value of the cost function starting from a poor
sample.

\begin{acknowledgments}

We would like to thank F.~Hamze, M.~Hernandez, J.~Oberoi, P.~Ronagh,
G.~Rosenberg, and Z.~Zhu for useful discussions.  A.~J.~O.~, D.~C.~J.~, and H.~G.~K.~ acknowledge
support from the NSF (Grant No.~DMR-1151387).  This research was based
upon work supported in part by the Office of the Director of National
Intelligence (ODNI), Intelligence Advanced Research Projects Activity
(IARPA), via MIT Lincoln Laboratory Air Force Contract
No.~FA8721-05-C-0002.  The views and conclusions contained herein are
those of the authors and should not be interpreted as necessarily
representing the official policies or endorsements, either expressed or
implied, of ODNI, IARPA, or the U.S.~Government.  The U.S.~Government is
authorized to reproduce and distribute reprints for Governmental
purpose.

\end{acknowledgments}

\bibliography{refs,comments}

\begin{thebibliography}{58}
\bibitem[{com({\natexlab{a}})}]{comment:bias}
\bibinfo{note}{Studies on different generations of the D-Wave quantum annealer
  \cite{boixo:13a,albash:15a,king:16} have suggested that the sampling might be
  biased with the first systematic study presented in Ref.~\cite{mandra:17}.}
\bibitem[{com({\natexlab{b}})}]{comment:dorband}
\bibinfo{note}{This greedy-descent type method was recently proposed by Dorband
  \cite{dorband:18x}.}
\bibitem[{com({\natexlab{c}})}]{comment:cpu}
\bibinfo{note}{In our studies we have typically used between $2^{17}$ and
  $2^{20}$ cluster updates per instance. This, in general, is more than is
  typically needed. However, we wanted to verify that the results do not change
  after running the simulations with many additional cluster updates. The
  typical numerical effort needed to perform one cluster update step depends on
  the size of the initial pool of solutions. For small pools, as used in the
  study of three-dimensional lattices, a simulational step (one cluster update)
  takes approximately $10.5\mu$s for $N = 64$ to $44.8\mu$s for $N = 216$ on an
  Intel Xeon E5-2680 v4 2.40GHz processor. In comparison, for a much larger
  solution pool, as used for the problems defined on chimera lattices that
  include excited states, each simulation step takes between $1570\mu$s for $N
  = 512$ to $3690\mu$s for $N = 968$. This, of course, can be improved with
  better memory management.}
\bibitem[{com({\natexlab{d}})}]{comment:params}
\bibinfo{note}{The parameters used to generate the data can be found in
  Ref.~\cite{mandra:17}.}
\bibitem[{com({\natexlab{e}})}]{comment:qubits}
\bibinfo{note}{Not all qubits on the D-Wave 2X device were operational. For a
  chimera lattice with nominally $8 \times c^2$ sites, the following number of
  qubits were addressable: for $c = 8$, $N = 483$ of $512$ qubits; for $c = 9$,
  $N = 615$ of $648$ qubits; for $c = 10$, $N = 758$ of $800$ qubits; for $c =
  11$, $N = 915$ of $968$ qubits; and for $c = 12$, $N = 1097$ of $1152$
  qubits.}
\end{thebibliography}

\begin{appendix}
\section{Limit distribution for uniformly distributed random numbers}
\label{app}

\begin{figure}[h]
\begin{centering}
\includegraphics[width=0.45\textwidth]{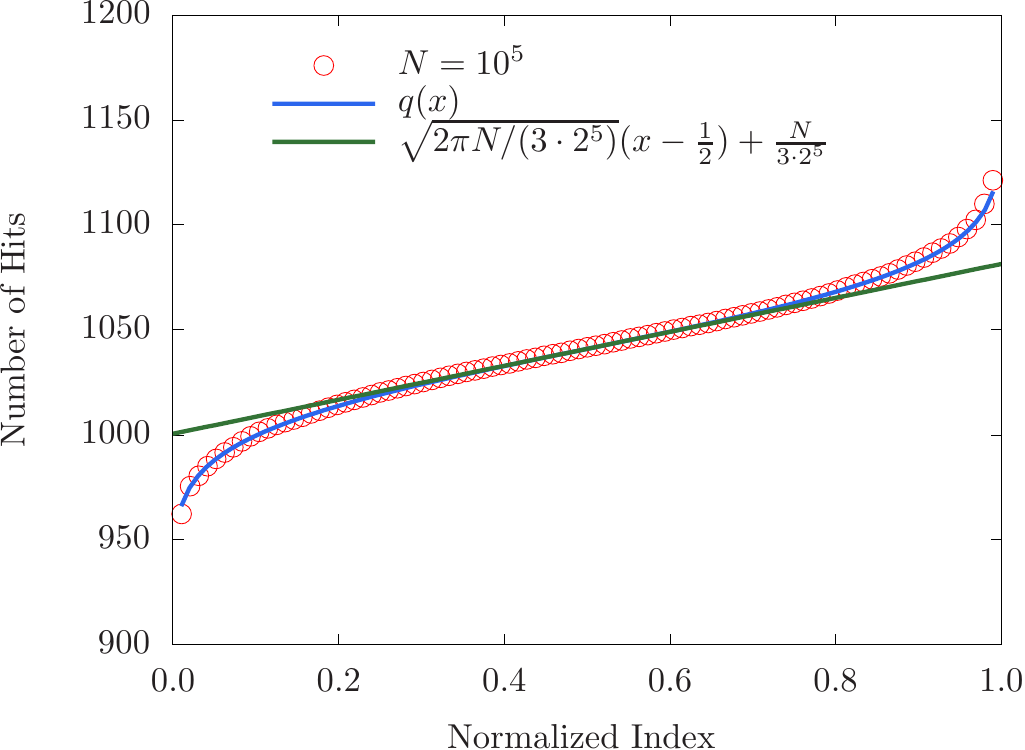}
\par\end{centering}
\caption{\label{fig:uniform1}Number of hits for $N=10^{5}$ uniformly distributed random numbers among $3\times2^{k}$
different values, with $k=5$. As expected, the histogram of number of hits follows
the inverse of the error function.}
\end{figure}

\begin{figure}[h]
\begin{centering}
\includegraphics[width=0.45\textwidth]{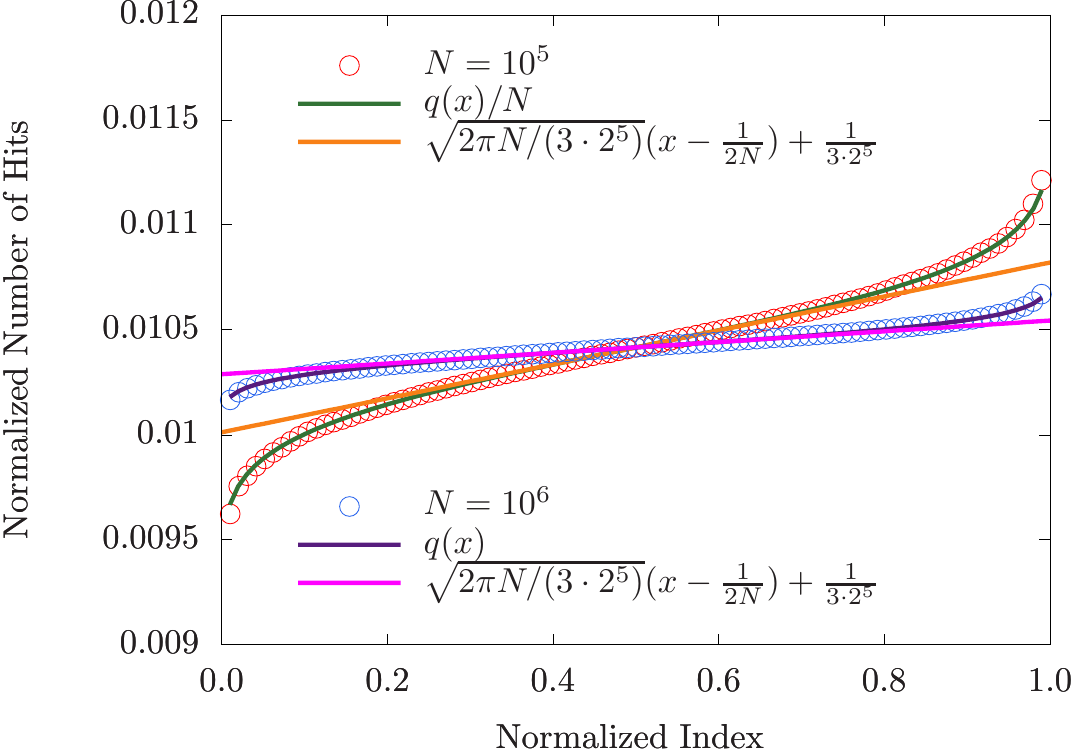}
\par\end{centering}
\caption{\label{fig:uniform2}Normalized number of hits for $N=10^{5}$ and
$N=10^{6}$ uniformly distributed random numbers among $3\times2^{k}$
different values, with $k=5$. As shown in the figure, the normalized histogram becomes flatter by increasing the number of random
numbers.}
\end{figure}

To understand the limit distribution of uniformly distributed random
numbers, let us call $N$ the number of uniformly sampled random numbers
among $M=3\times2^{k}$ possible values (which correspond to the possible
ground state configurations of a given instance). Since all the random
numbers are uniformly sampled, the probability for a given value $m$
to have a number of hits $q$ is a binomial distribution of the form
\begin{equation}
P_{m}(q)=\binom{N}{q}p^{q}(1-p)^{N-q},\label{eq:gs_limit}
\end{equation}
where $p=\frac{1}{M}$ is the probability that such a value is chosen
every time a random number is extracted. In the limit of large $N$
and assuming that all the $P_{m}(q)$ are independent random variables,
the histogram of the number of hits will follow a normal distribution
with $\mu=\frac{N}{M}$ and variance $\sigma^{2}=\frac{N}{M}\left(1-\frac{1}{M}\right)\approx\frac{N}{M},$ that
is, $q=\mathcal{N}(\mu,\sigma^{2})$. Therefore, after the reordering
of the indices following the number of hits, the histogram should
follow the inverse of the error function, that is,
\begin{equation}
q(x)=\sqrt{\frac{2N}{M}}\text{erf}^{-1}(2x-1)+\frac{N}{M},\label{eq:hist_limit}
\end{equation}
where $x$ is the normalized index. Figures \ref{fig:uniform1} and
\ref{fig:uniform2} compare the limit distribution of $N=10^{5}$ and $N=10^6$
random numbers uniformly sampled from $M=3\times2^{k}$, with $k=5$, distinct
values. As one can see, the empiric distribution is consistent with the limit
distribution in Eq.~(\ref{eq:hist_limit}). As expected, the slope of the
normalized histogram is proportional to $\sqrt{\frac{1}{NM}}$ and it decreases
by increasing the number of samplings $N$.


\end{appendix}

\end{document}